\documentclass[a4paper,10pt, twocolumn]{article}
\usepackage{lineno,hyperref}
\usepackage{graphicx,graphics}
\usepackage{amsmath}
\usepackage{amssymb}
\usepackage{epstopdf}
\usepackage{amstext}
\usepackage{amsthm}
\usepackage{amsfonts}
\usepackage{latexsym}
\usepackage{array}
\usepackage{xfrac}
\usepackage{color}
\usepackage{multirow}
\usepackage{fontenc}
\usepackage{bm}
\usepackage{adjustbox}
\usepackage{dcolumn}
\usepackage{color}
\usepackage{subcaption}
\usepackage{graphicx}
\usepackage{multicol}
\begin{document}

\title{\Large\bf Comment on the manuscript 1806.02080v1 entitled ``Spurious finite-size instabilities of a 
new Gogny interaction suitable for astrophysical applications"}

\author{C. Gonzalez-Boquera, M. Centelles, X. Vi\~nas \\ \small Departament de F\'isica Qu\`antica i Astrof\'isica 
and Institut de Ci\`encies del Cosmos (ICCUB), \\ \small 
Facultat de F\'isica, Universitat de Barcelona, Mart\'i i Franqu\`es 1, E-08028 Barcelona, Spain
\and L.M. Robledo \\ \small Departamento de F\'isica Te\'orica,
Facultad de F\'isica, Universidad Aut\'onoma de Madrid,
\\ \small E-28049 Madrid, and Center for Computational Simulation,
Universidad Polit\'ecnica de Madrid, \\ \small 
Campus de Montegancedo, Boadilla del Monte, E-28660 Madrid, Spain}
%\affiliation{Aff2}
%\date{\small \today}
\date{}

    \twocolumn[
  \begin{@twocolumnfalse}
  \vspace*{-1cm}    \maketitle
\begin{abstract}
The conclusions of the manuscript 1806.02080v1 questioning the adequacy of 
the recently proposed Gogny D1M* interaction for finite nuclei calculations using harmonic
oscillator (HO) basis are revised. Several convergence and stability studies
are performed with HO basis of different sizes and oscillator parameters and
the results show the robustness of the D1M* results for finite nuclei. This 
analysis is also extended to beyond mean-field calculations of generator-coordinate-method
type with D1M*. On the other hand, the existence of a finite-size instability
in finite nuclei when coordinate space methods are used to solve the HF equations (as
shown in 1806.02080v1) is independently confirmed for D1M* using an
in-house computer code based on a quasilocal approximation to the HF
exchange potential. We confirm that the most affected quantity in the coordinate 
space calculation is the spatial density at the origin, but integrated
quantities like binding energies or radii show a plateau against the number of 
iterations, where they are consistent with the values from the HO basis calculation,
before diverging for a larger number of iterations. A connection between the last 
occupied $s$-orbital in the nucleus and the appearance of instabilities in 
coordinate space is observed.
\end{abstract}
%\linenumbers
  \end{@twocolumnfalse}
  ]
%\section{Introduction} 
In Ref.~\cite{gonzalez18} we proposed a new parametrization D1M* of the Gogny interaction, aimed to 
predict a stiffer equation of state of neutron-star matter and to reproduce maximum 
neutron star masses of $2 M_\odot$, in agreement with recent 
astrophysical observations \cite{demorest10, antoniadis13}. This property is not
achieved by the standard Gogny forces of the D1 family \cite{gonzalez17}.
We also wanted to preserve the good description of finite 
nuclei at the Hartree-Fock-Bogoliubov (HFB) level provided by the D1M force \cite{goriely09}. 
In the fit of D1M* \cite{gonzalez18}, we modified the eight 
finite-range strengths of the D1M force keeping the other parameters at their
D1M values. Seven linear combinations of these strengths, related to 
different properties of symmetric nuclear matter, and the pairing strengths in the
$S$=0, $T$=1 channel, were constrained to take the same values as in  D1M. 
The eighth combination was used to modify the slope of the
symmetry energy and, therefore, the prediction for the maximum neutron star mass. Finally, the 
$t_3$ parameter was fine tuned to obtain a finite nuclei description with 
similar quality to D1M for the same set of nuclei.
All of the finite nuclei calculations in \cite{gonzalez18} were carried out with the \mbox{HFBaxial} code \cite{robledo02} 
using an approximate second-order gradient method to solve the HFB 
equations in a harmonic oscillator (HO) basis including up to 19 major oscillator
shells and the oscillator lengths adapted to the characteristic length-scale 
dependence with mass number. Notice that HFB calculations of deformed nuclei
with Gogny interactions are usually performed in a HO basis since the seminal paper
 of Decharg\'e and Gogny \cite{decharge80}.    

In the recent manuscript 1806.02080v1 \cite{martini18}, it is found that 
this new D1M* Gogny force and the D1N force suffer from
the existence of spurious finite-size instabilities in the $S$=0, $T$=1 
channel. These instabilities are detected through a fully 
antisymmetrized RPA calculation of the nuclear matter response 
functions based on the continued fraction technique \cite{depace16}. 
This procedure was applied in \cite{depace16} to the search of instabilities 
in standard Gogny forces as well as forces of Gogny type including tensor terms.
In this study, in agreement with the results of 
similar analyses carried out for Skyrme functionals \cite{hellemans13}, it was concluded that 
the key quantity to detect spurious finite-size instabilities is the critical density, 
which makes the nucleus unstable if it reaches a value $\rho_c \simeq 
0.20$ fm$^{-3}$ for a momentum transfer of about 2.5 fm$^{-1}$. This 
critical density of $\rho_c \simeq 0.20$ fm$^{-3}$ may be reached in HF 
calculations of some nuclei, as for example $^{40}$Ca. 
The instabilities of D1M* and D1N were predicted in nuclear matter \cite{martini18,depace16} and 
their appearance in calculations of nuclei in coordinate space is verified in \cite{martini18} by 
performing HFB calculations of finite nuclei on a mesh 
assuming spherical symmetry, using the Gogny forces with the FINRES$_4$ 
code \cite{bennaceur18}. From the results shown in 
\cite{martini18}, it can be seen that the neutron and proton 
density profiles are very delicate quantities, which are largely 
affected in the center of the nucleus by the finite-size instabilities, 
showing a continuously increasing or decreasing trend as a function of 
the number of iterations in the iterative solution of the non-linear HFB equation, without reaching a stabilized value.
In another paper \cite{hellemans13}, a similar study was done in the framework of Skyrme forces.

In the present manuscript we provide additional information 
about the possible impact of the finite-size instability in the $S=0$, $T=1$ channel,
detected in \cite{martini18}, on the binding energies,
neutron and proton radii and density profiles of finite nuclei computed using the D1M* 
interaction with the HO basis \cite{gonzalez18}. The tests performed support the
fact that the predictions for finite nuclei obtained with D1M* with the HO basis 
in~\cite{gonzalez18} are robust.
We also perform spherical HF calculations with D1M* on a mesh in coordinate space.
In spite of the non-convergent behavior of the nucleon density profiles,
there is an optimal number of iterations for which integrated quantities such as the 
total binding energies present a plateau pattern.
When we compare the D1M* binding energies obtained with the HO basis  
and these values computed on a mesh, we observe an excellent agreement between both sets of results.
The quality of this agreement in D1M* is equivalent to the one we find in the same type of comparison 
if we use D1M, which does not have finite-size instabilities.

\begin{figure}[h!]
\vspace*{2.3cm}
\includegraphics[width=1.\columnwidth,clip=true]{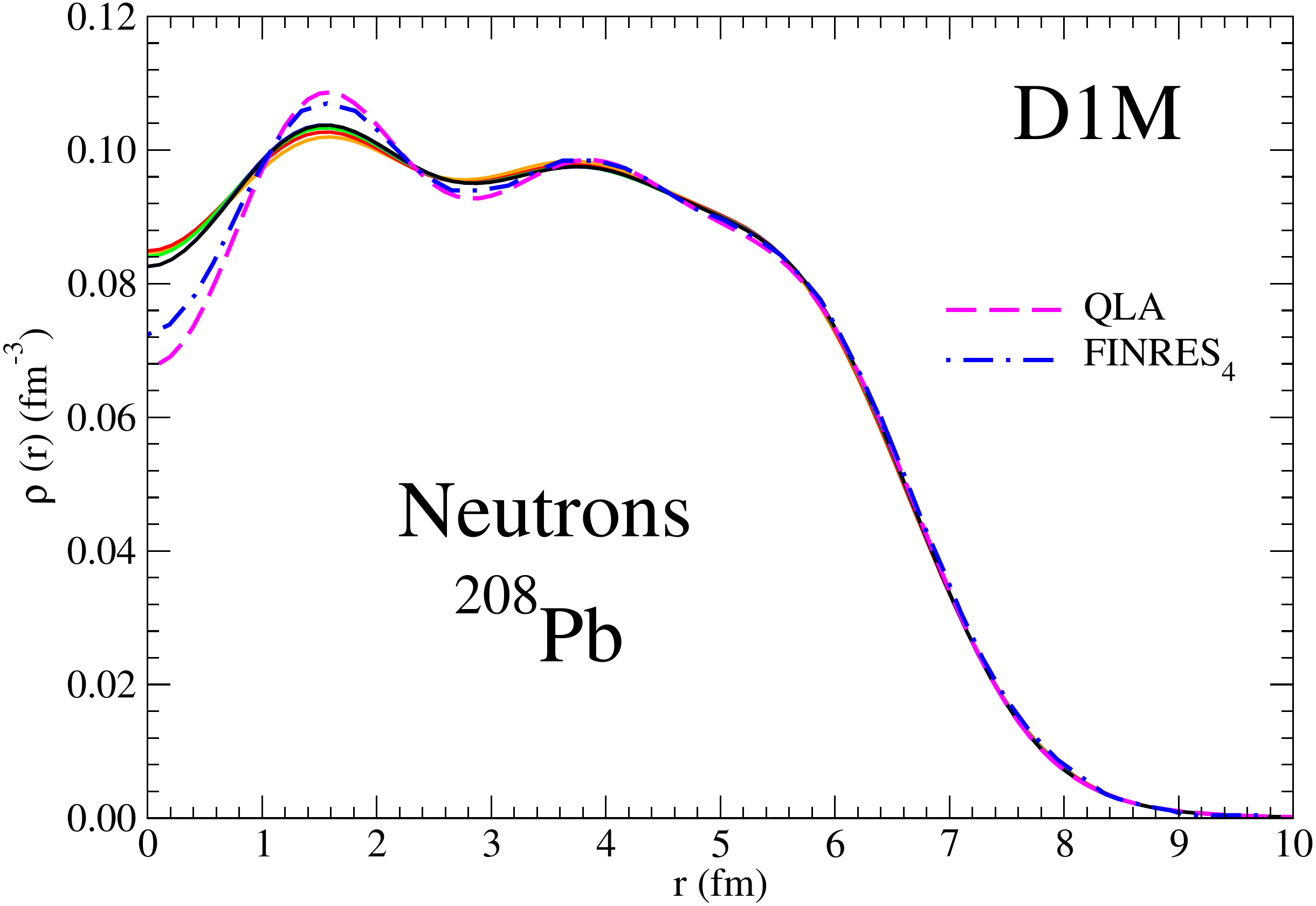}
\caption{Solid lines: Neutron density of $^{208}$Pb computed with the D1M force
with a HO basis with 12, 14, 16, 18 and 19 shells. Dashed lines: 
The same density obtained through a HF calculation on a mesh in the quasilocal
approach. Dash-dotted lines: The same density extracted from Fig. 3 in Ref.~\cite{martini18}.}\label{fig:neutronessD1M}
\end{figure}

\begin{figure}[h!]
\vspace*{2.4cm}
\includegraphics[width=1.\columnwidth,clip=true]{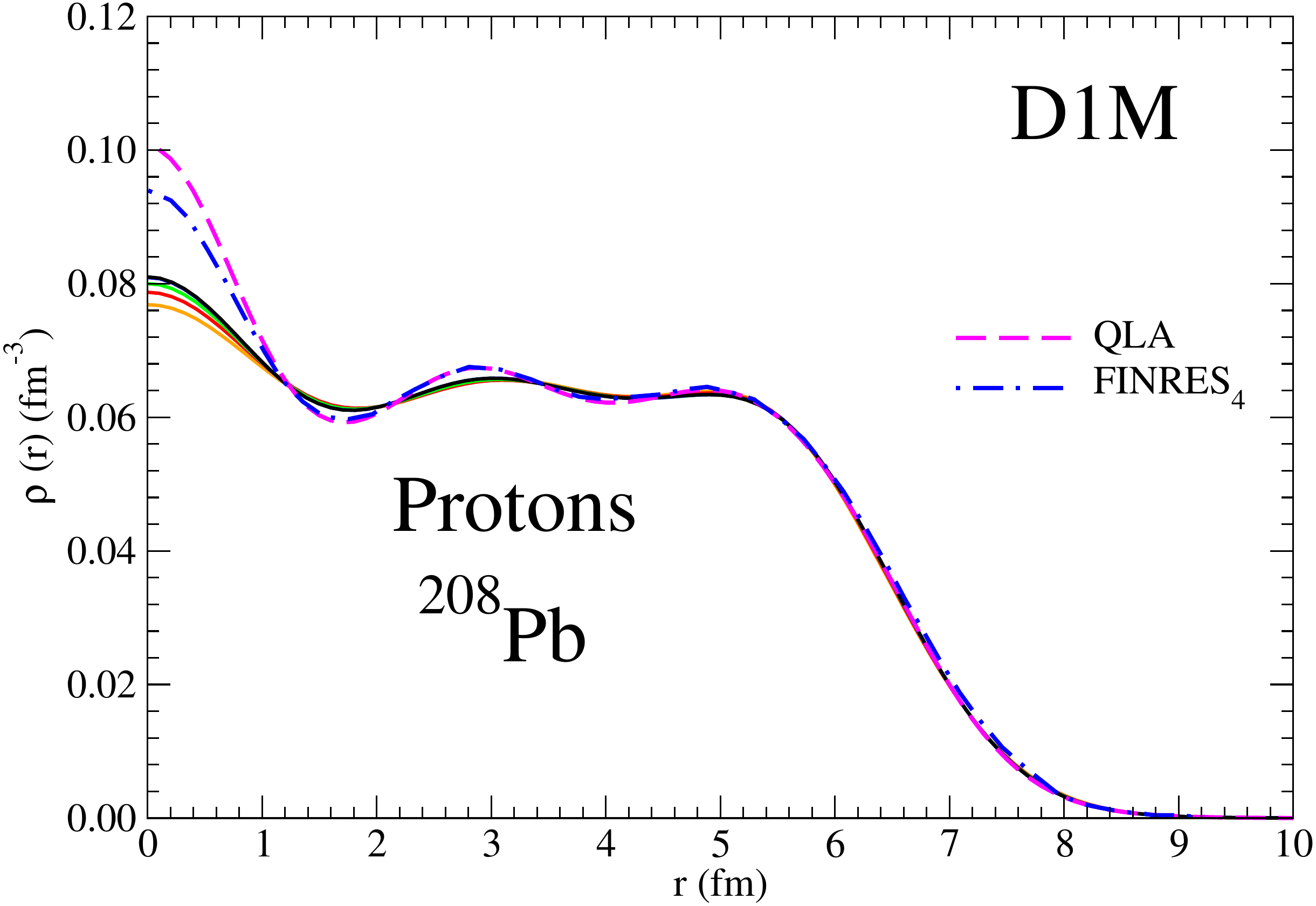}
\caption{The same as in Fig.~\ref{fig:neutronessD1M} but for the proton density.}\label{fig:protonesD1M}
\end{figure}

\begin{figure}[h!]
\includegraphics[width=1.\columnwidth,clip=true]{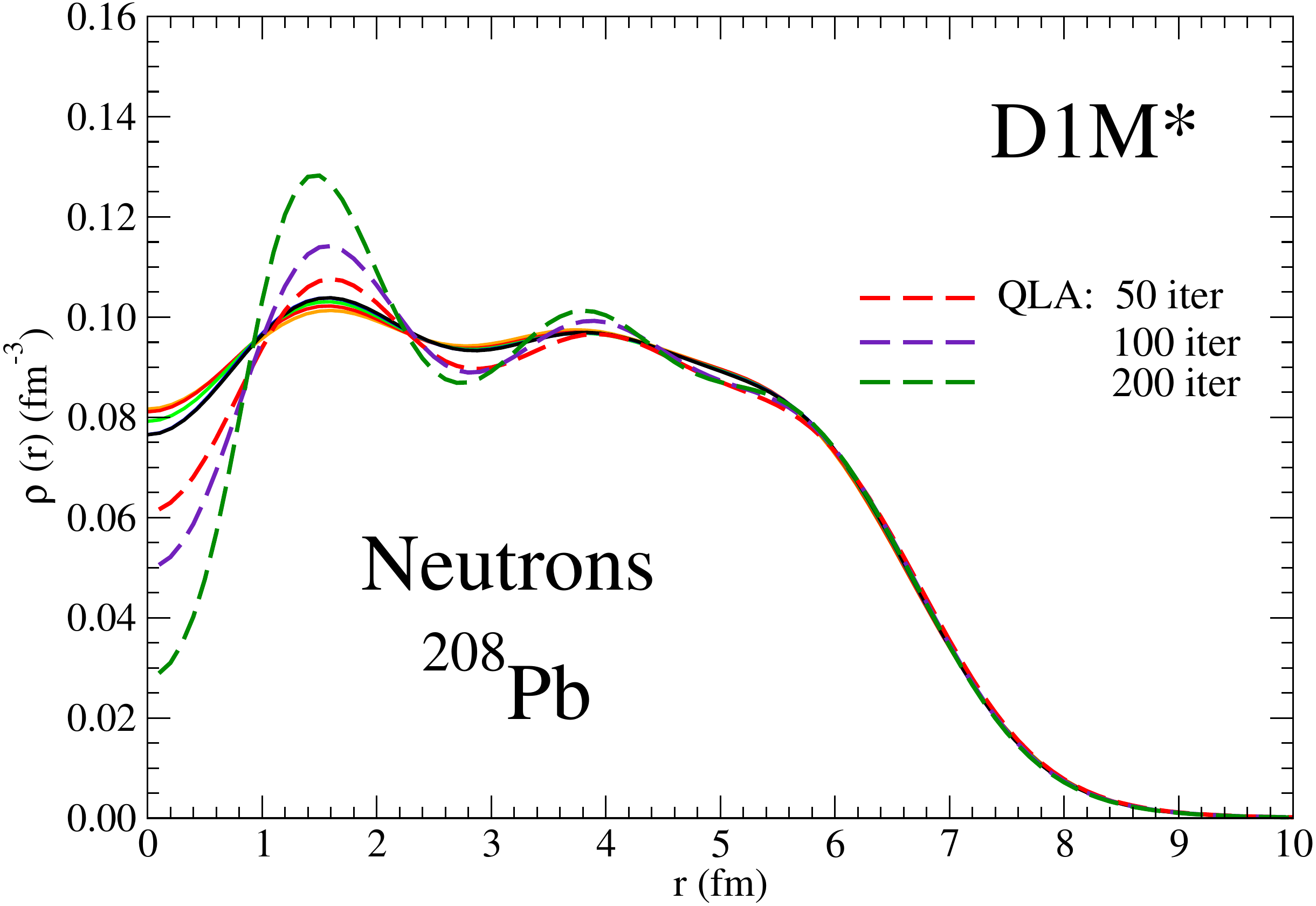}
\caption{The same as in Fig.~\ref{fig:neutronessD1M} but computed with the D1M* force.
The HF density is displayed for three different number of iterations.}\label{fig:neutronesD1Ms}
\end{figure}

\begin{figure}[h!]
\includegraphics[width=1.\columnwidth,clip=true]{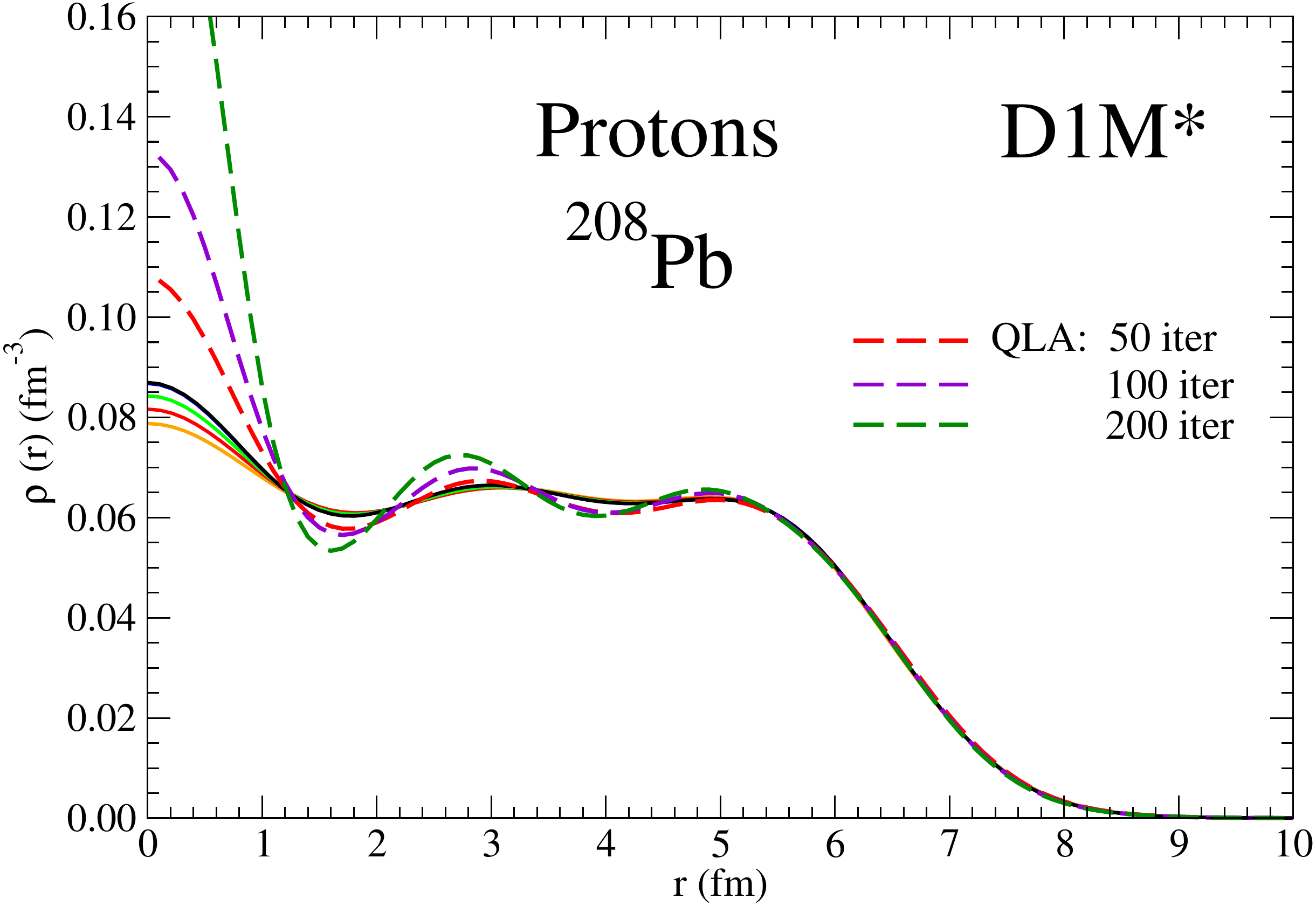}
\caption{The same as in Fig.~\ref{fig:neutronesD1Ms} but for the proton density.}\label{fig:protonesD1Ms}
\end{figure}

In our finite-nuclei deformed calculations performed using the HO basis for over 600 even-even 
nuclei and covering the whole nuclear chart \cite{gonzalez18}, we did not face
any lack of convergence in the nucleon density profiles similar to the one 
discussed in \cite{martini18}. The difference between the two calculations 
is our use of a HO basis instead of using a mesh as in Ref.~\cite{martini18}. 
It is well known that the HO basis introduces an ultraviolet cutoff,
due to its Gaussian falloff in momentum space, which acts as a regulator 
for the behavior related to high-momentum components in the wave function. 
On the other hand, mesh calculations are more prone to suffer the effect of possible
ultraviolet divergences. This difference between HO basis and mesh calculations 
was already recognized in \cite{martini18}, where it was argued that the use of a 
HO basis ``strongly renormalizes the interaction and inhibits
the development of instabilities" and that ``the D1M* interaction should 
only be used with the basis employed to fit its parameters". 
From this statement one should expect significant changes in the value of physical 
observables computed with the D1M* force when the HO basis size is increased 
or its harmonic oscillator lengths modified. 
To test this statement, we have carried out calculations with different
HO basis sizes including 11, 13, 15, 17, 19 and 20 full HO shells for some
representative nuclei using both D1M* and D1M. Larger basis are impractical
for Gogny calculations involving deformed nuclei and therefore are not currently
implemented in the HFBaxial code. In all the cases analyzed,
we observe a similar behavior of the observables as a function of the number
of shells in both the D1M* and D1M cases. The range of nuclei considered
includes deformed nuclei like $^{224}$Ra, $^{168}$Er or $^{48}$Cr 
and spherical nuclei like $^{16}$O, $^{40}$Ca, $^{56}$Ni, $^{100}$Sn, $^{132}$Sn 
or $^{208}$Pb. Except
for the binding energy (which is the variational magnitude and therefore always
increases with increasing basis size), the changes in the other observables
(radii, quadrupole deformation, octupole deformation, etc.) are less than
one in a thousand when going from the smallest to the largest basis size. 
Interestingly, the convergence rate with basis size of the
density at the origin is rather slow and requires a large number of shells
both in D1M and D1M*, and even in the D1S case, as can be seen in 
Figs.~\ref{fig:neutronessD1M}--\ref{fig:protonesD1Ms},
%\ref{fig:neutronesD1M,fig:protonesD1M,fig:neutronesD1Ms,fig:protonesD1Ms}
where the neutron and proton densities of $^{208}$Pb computed with
D1M and D1M* with different number of HO shells are displayed by solid lines.  
It is to be pointed out that the central density does not 
enter significantly in most of the observables like radii or multipole
moments as the corresponding operators go to zero at the origin. Also 
the energy, which should be more sensitive through the strongly repulsive
density-dependent part of the interaction to the slow converge rate of the central density, shows
a smooth behavior. \nolinebreak  
We also would like to mention that in Ref.~\cite{PRC89.054310} we
studied fission properties of the uranium isotopes including very neutron-rich 
isotopes using the parametrizations D1S, D1M and D1N.
From some of the claims of Ref.~\cite{martini18} questioning the fact that one
usually uses different HO parameter values for different nuclei and shapes,
one could have expected to observe erratic results in the calculations with 
D1N, which may also show finite-size instabilities~\cite{martini18}. 
However this is not the case, 
in spite of using in the fission calculations, which require very large 
deformations, a HO basis (size and oscillator lengths) very different 
from the one used in the D1N fit, tailored to the ground state.
This is an additional indication that the regularization of
the ultraviolet sector of the force as a consequence of using a HO basis renders
the predictions of D1N as reasonable as the ones of D1S and D1M.

In order to analyze more deeply the findings of Ref.~\cite{martini18} concerning finite nuclei 
calculations on a mesh with D1M*, we have performed 
HF calculations using the method described in \cite{soubbotin03}, where the HF exchange energy
is approximated by a quasilocal approach calculated using the
extended Thomas-Fermi expansion of the density matrix derived in 
\cite{soubbotin00}. This approach is similar to the density matrix 
expansions proposed by Negele and Vautherin \cite{negele72} or by Campi 
and Bouyssy \cite{campi78}. The quasilocal approximation allows one to 
write the energy density functional for finite-range effective interactions
in a local form and enables HF calculations in 
coordinate space in a similar fashion to the case of Skyrme forces \cite{vautherin72}. 
This quasilocal approach, which is discussed in detail in \cite{soubbotin03}, gives results that are 
very close to the full HF results calculated with HO 
basis as can be seen in Refs.~\cite{soubbotin03,krewald06,behera16}. 
The HF nucleon densities we obtain in a mesh with the quasilocal approximation 
nicely agree with the corresponding densities obtained using the FINRES$_4$ code 
\cite{bennaceur18}, as can be seen from Figs.~\ref{fig:neutronessD1M} and 
\ref{fig:protonesD1M} for the $^{208}$Pb neutron and proton densities
computed with the D1M interaction. From these figures we also see that the 
shell oscillations in the center of the nucleus of the densities obtained
by the mesh calculation are more pronounced than the shell oscillations 
provided by the HO basis. This is a first qualitative indication that the
mesh densities may be more affected by the finite-size instabilities than
the densities calculated with a HO basis.  
We shall also point out that the quasilocal approach is very well 
suited for the analysis of instabilities in spherical nuclei in coordinate 
space, as it reproduces all the instabilities of finite nuclei 
reported in \cite{martini18}.

\begin{table}[t]
\centering
\caption{Quantal D1M HF binding energies in doubly magic nuclei
computed using the HO basis and the quasilocal approximation in
coordinate space (mixing factor 0.9).}
%\\
%\renewcommand{\tabcolsep}{0.45cm}
%\renewcommand{\arraystretch}{1.2}
%\resizebox{\columnwidth}{!}{ 
\begin{tabular}{ccc}
\hline
\hline
\small Nucleus &\small $B_{\rm HO}$(MeV) &\small $B_{\rm QLA}$(MeV)  \\
\hline
\hline
$^{16}$O   & 128.02 & 127.02\\
$^{40}$Ca  & 341.67 & 340.53\\
$^{56}$Ni  & 478.65 & 478.17\\
$^{90}$Zr  & 781.26 & 781.29\\
$^{132}$Sn &1102.57 &1103.31\\
$^{208}$Pb &1636.08 &1637.96\\
\hline
\end{tabular}
%}
\label{table1}
\end{table}

\begin{table}[t]
\centering
\caption{Quantal D1M* HF binding energies in doubly magic nuclei
computed using the HO basis and the quasilocal approximation
in coordinate space with 100 and 150 iterations (mixing factor 0.9).}
%\\
%\renewcommand{\tabcolsep}{0.45cm}
%\renewcommand{\arraystretch}{1.2}
%\resizebox{\columnwidth}{!}{  
\begin{tabular}{cccc}
\hline
\hline
\small Nucleus &\small $B_{\rm HO}$(MeV) &\small $B_{\rm QLA}^{100}$(MeV) &\small $B_{\rm QLA}^{150}$(MeV) \\
\hline
\hline
$^{16}$O   & 128.32  & 127.29 & 127.29\\
$^{40}$Ca  & 342.55  & 341.33 & 341.35\\
$^{56}$Ni  & 481.34  & 479.41 & 479.41\\
$^{90}$Zr  & 783.14  & 782.58 & 782.60\\
$^{132}$Sn & 1103.08 &1103.25 &1103.82\\
$^{208}$Pb & 1636.73 &1637.83 &1638.18\\
\hline
\end{tabular}
%}
\label{table2}
\end{table}

To test the consistency between the HO basis and mesh quasilocal calculations, we show 
in Table~\ref{table1} the binding energies computed with the D1M force in a HO basis 
using the HFBaxial code for the magic nuclei $^{16}$O, $^{40}$Ca, $^{56}$Ni, $^{90}$Zr 
(all of them with 15 HO shells), $^{132}$Sn (with 17 HO shells) and $^{208}$Pb (with 19 
HO shells). The oscillator lengths have been optimized as to minimize the energy. In the 
same table we display the D1M binding energies from the spherical HF calculation on a mesh in 
coordinate space using the quasilocal approach, which are well converged as far as the 
D1M force is free from finite-size instabilities, in agreement with \cite{martini18}. It 
can be seen that the results from the HF method in the HO basis and on a mesh are in 
excellent agreement among them. Notice that the two solutions cannot be identical 
because, besides the fact that the phase spaces and numerical methods of the two 
calculations are very different, the mesh calculation implements the quasilocal 
approximation of the exchange potential. In Table~\ref{table2} we show the binding 
energies predicted by the D1M* force for the same nuclei computed with both the HO basis 
and the quasilocal approach in coordinate space after certain numbers of iterations. We 
can see that the D1M* HF quasilocal energies show a nearly stable behavior over a range 
of iterations. In this range the agreement between the D1M* binding energies computed in 
a HO basis and in a mesh is similar to that found using the D1M force. The binding 
energies reported in Tables~\ref{table1} and~\ref{table2} show the consistency of the 
HFB results in a HO basis, which do not depend on details such as the size and 
oscillator lengths of the basis within the range of 20 oscillator shells. The impact of 
the finite-size instabilities in the D1M* calculations on a mesh is much more relevant 
in the neutron and proton density profiles around the center of the nucleus as can be 
seen in Figs.~\ref{fig:neutronesD1Ms} and \ref{fig:protonesD1Ms} for $^{208}$Pb. The 
nucleon densities from the mesh calculation in the quasilocal approach with D1M* 
exhibit a divergent trend against the number of iterations, which does not happen when 
the densities are calculated with the D1M force. This situation with the HF densities in 
coordinate space is different from the one shown by the densities computed in a HO 
basis, where the central densities exhibit a convergent trend with the number of HO 
shells used in the calculation for both the D1M and D1M* forces.

\begin{figure}[h!]
 \centering
 \includegraphics[width=0.95\linewidth, clip=true]{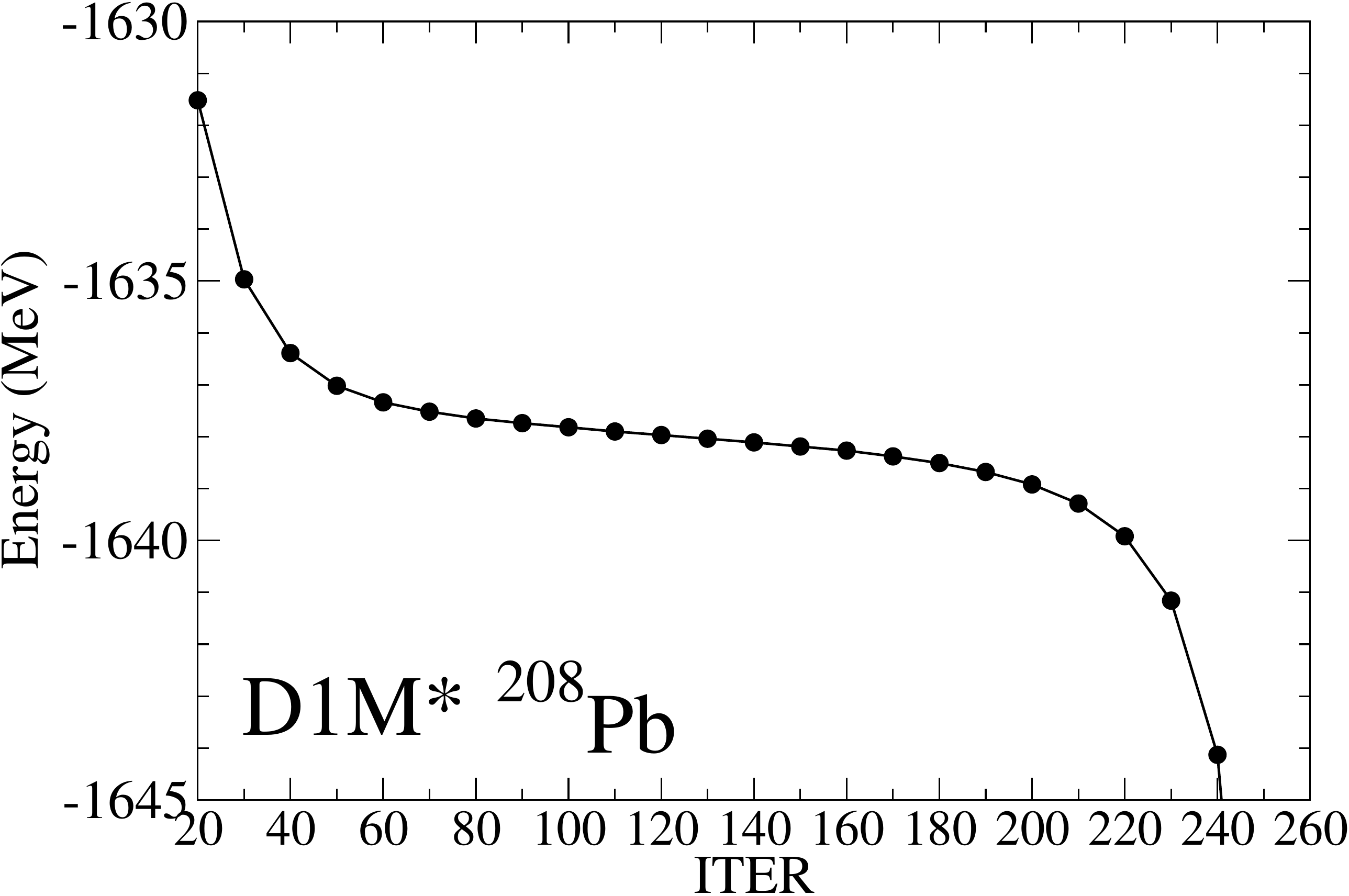}
  \caption{Total energy against the number of iterations of the calculation on a mesh, for $^{208}$Pb with the D1M$^{*}$ interaction.}\label{fig:208Pb_Energy}
\end{figure}

\begin{figure}[h]
 \centering
 \includegraphics[width=0.95\linewidth, clip=true]{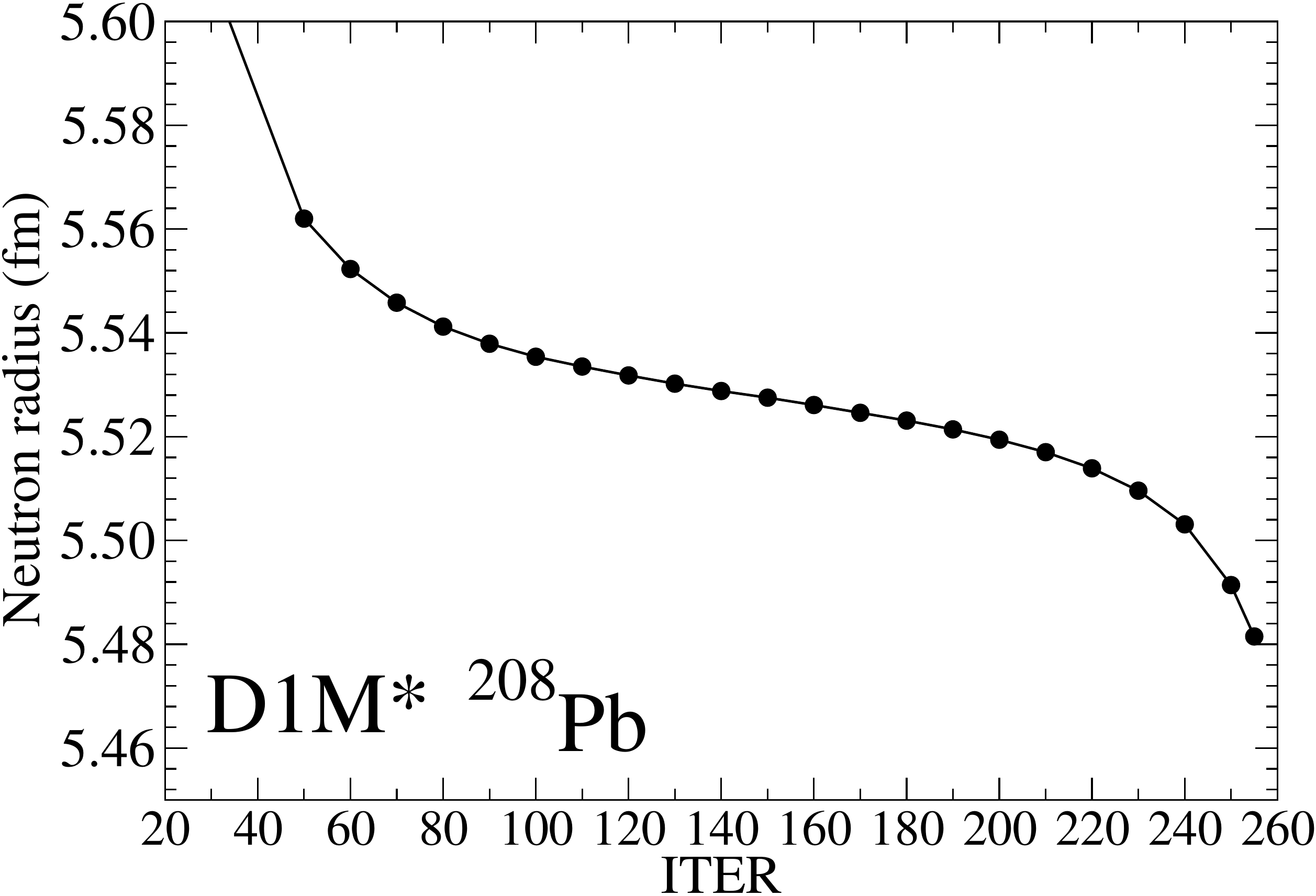}
  \caption{Neutron radius against the number of iterations of the calculation on a mesh, for $^{208}$Pb with the D1M$^{*}$ interaction.}\label{fig:208Pb_rn}
\end{figure}

\begin{figure}[b!]
 \centering
 \includegraphics[width=0.95\linewidth, clip=true]{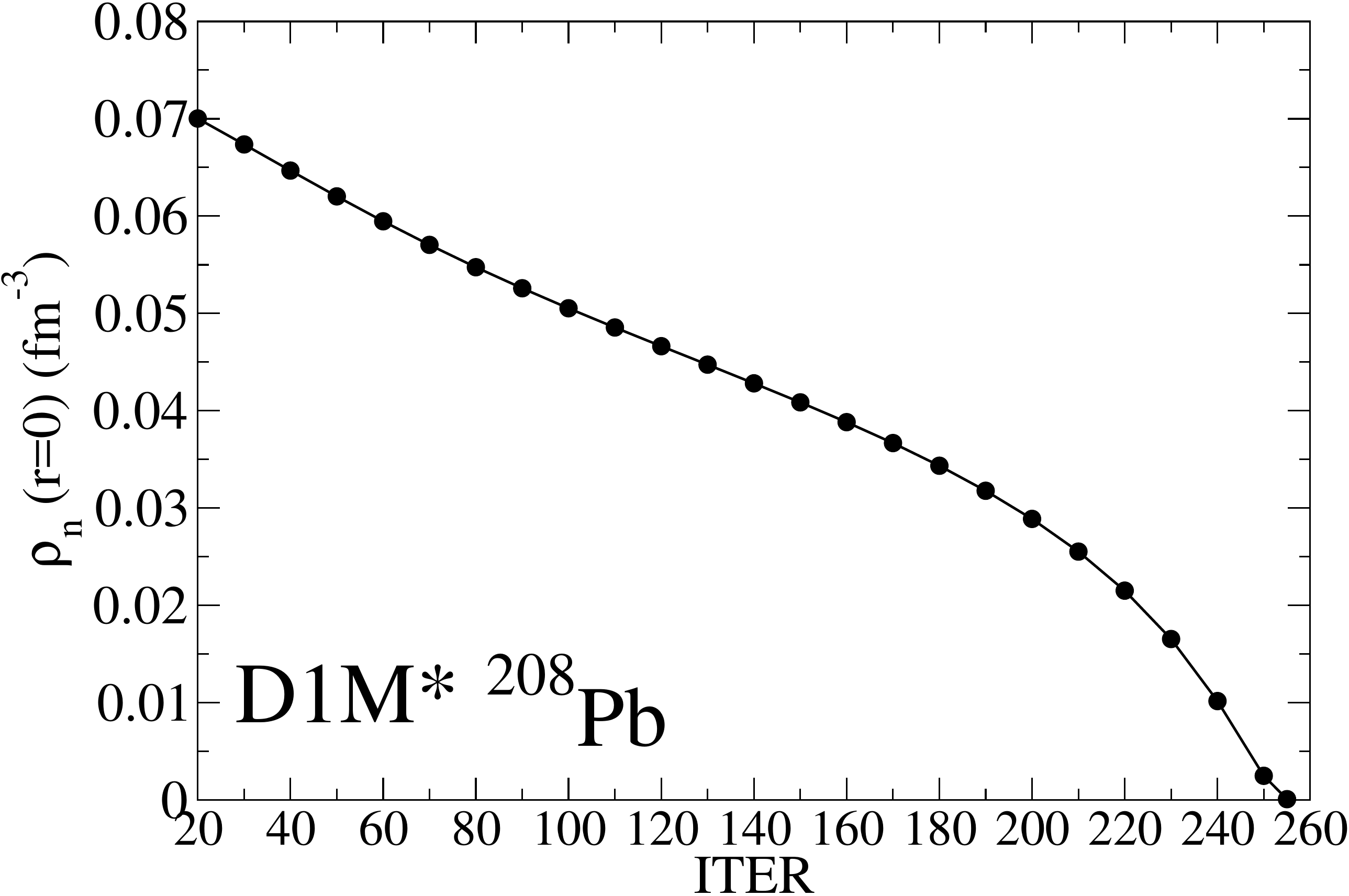}
  \caption{Neutron density at the origin against the number of iterations of the calculation on a mesh, for $^{208}$Pb with the D1M$^{*}$ interaction.}\label{fig:208Pb_rhon_ol}
\end{figure}

\begin{figure}[t!]
 \centering
 \vspace{1.1cm}
 \includegraphics[width=0.95\linewidth, clip=true]{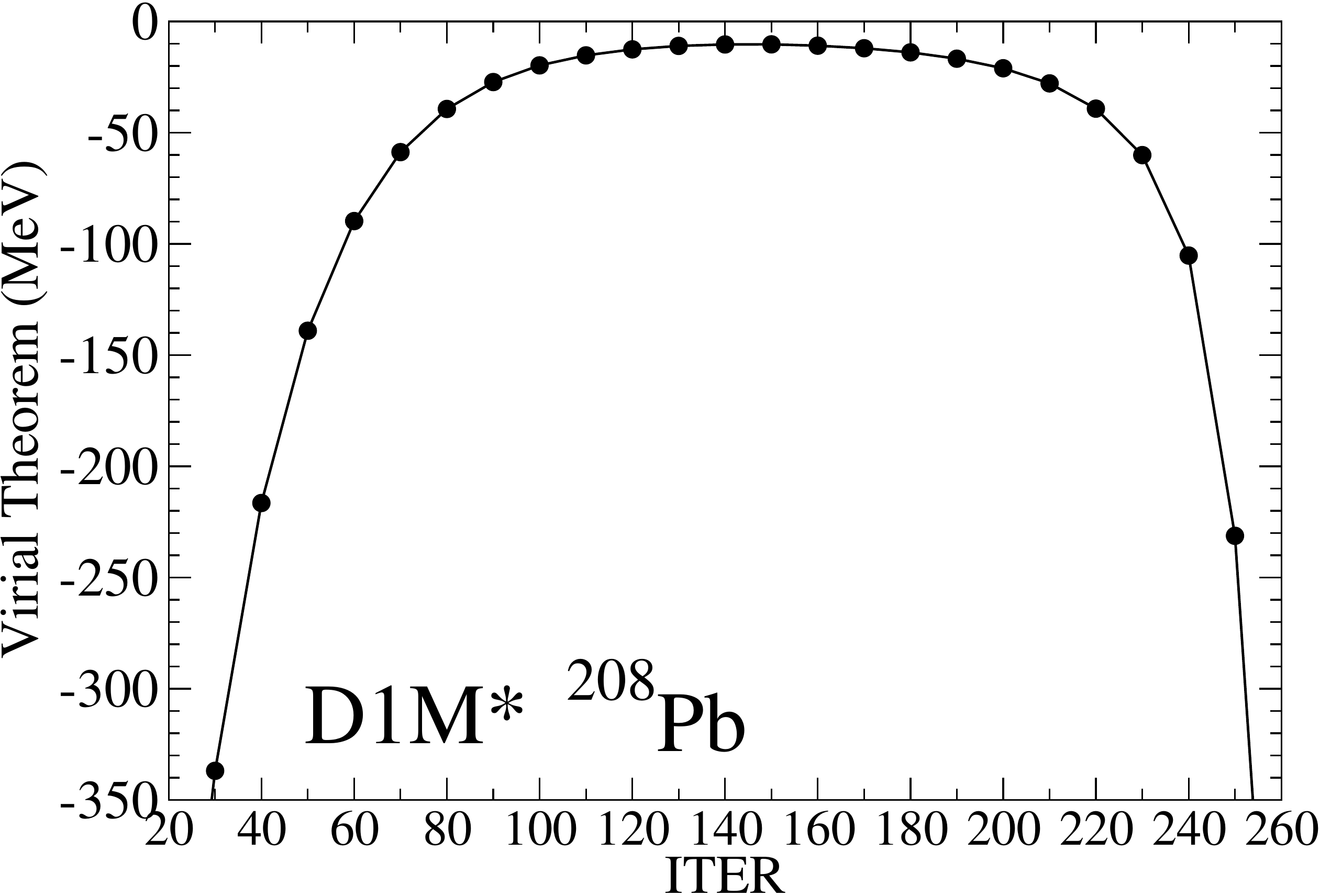}
  \caption{Virial theorem against the number of iterations of the calculation on a mesh, for $^{208}$Pb with the D1M$^{*}$ interaction.}\label{fig:208Pb_Virial}
\end{figure}

\begin{figure}[h!]
 \centering
  \vspace{0.2cm}
 \includegraphics[width=0.95\linewidth, clip=true]{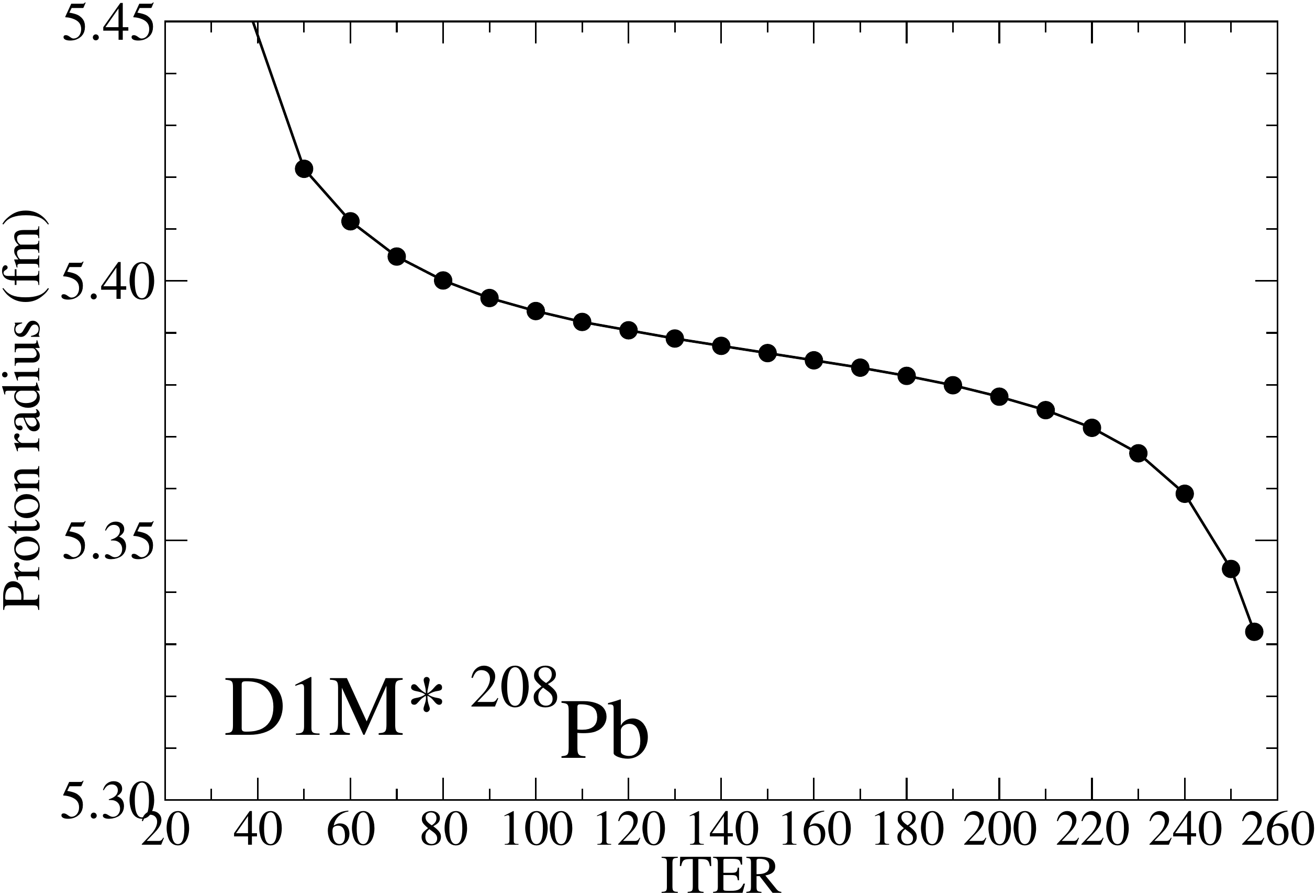}
  \caption{Proton radius against the number of iterations of the calculation on a mesh, for $^{208}$Pb with the D1M$^{*}$ interaction.}\label{fig:208Pb_rp}
\end{figure}

\begin{figure}[b!]
 \centering
 \vspace*{0.11cm}
 \includegraphics[width=0.95\linewidth, clip=true]{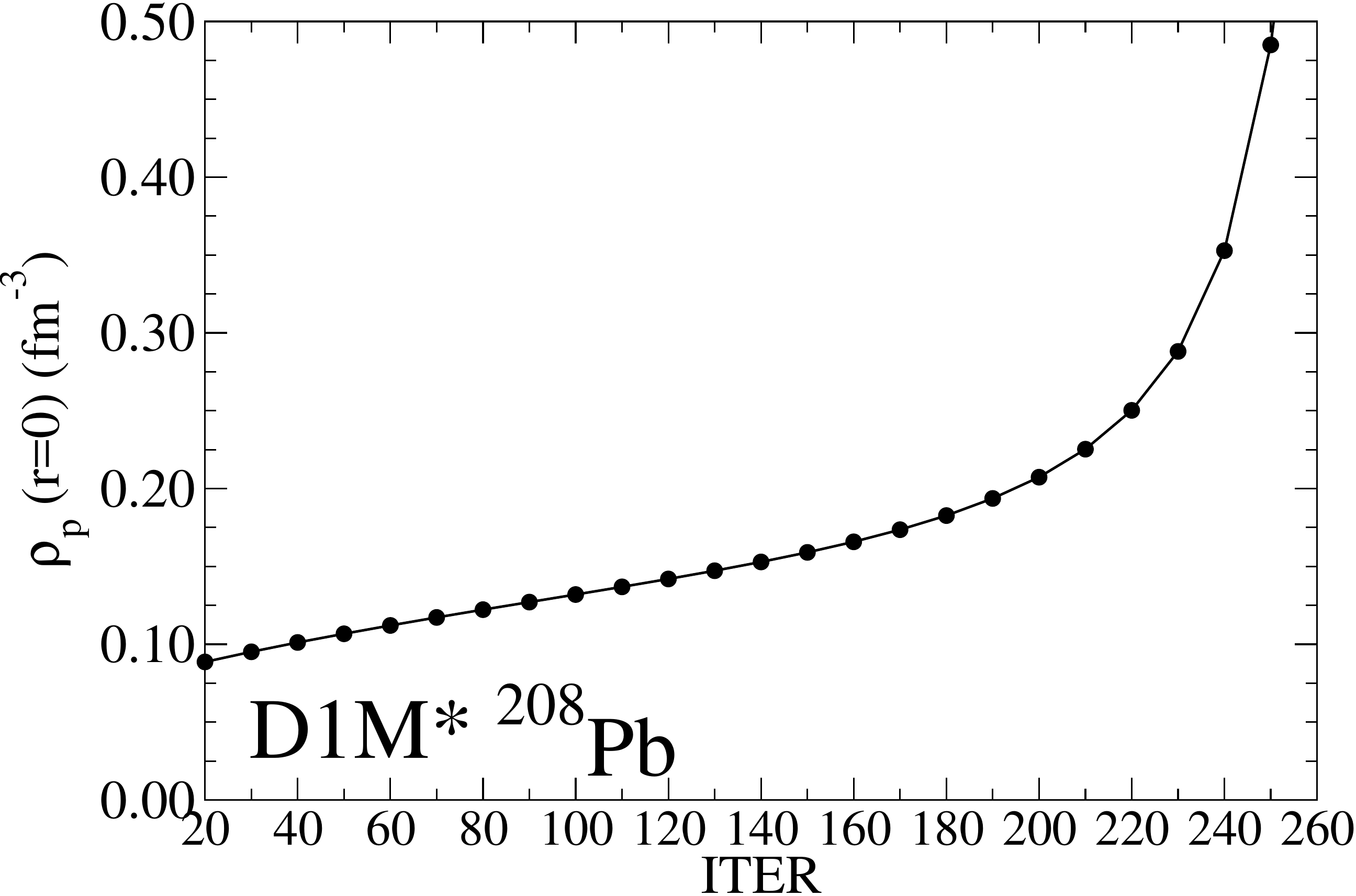}
  \caption{Proton density at the origin against the number of iterations of the calculation on a mesh, for $^{208}$Pb with the D1M$^{*}$ interaction.}\label{fig:208Pb_rhop_o}
\end{figure}

\begin{figure}[h]
 \centering
 \includegraphics[width=0.95\linewidth, clip=true]{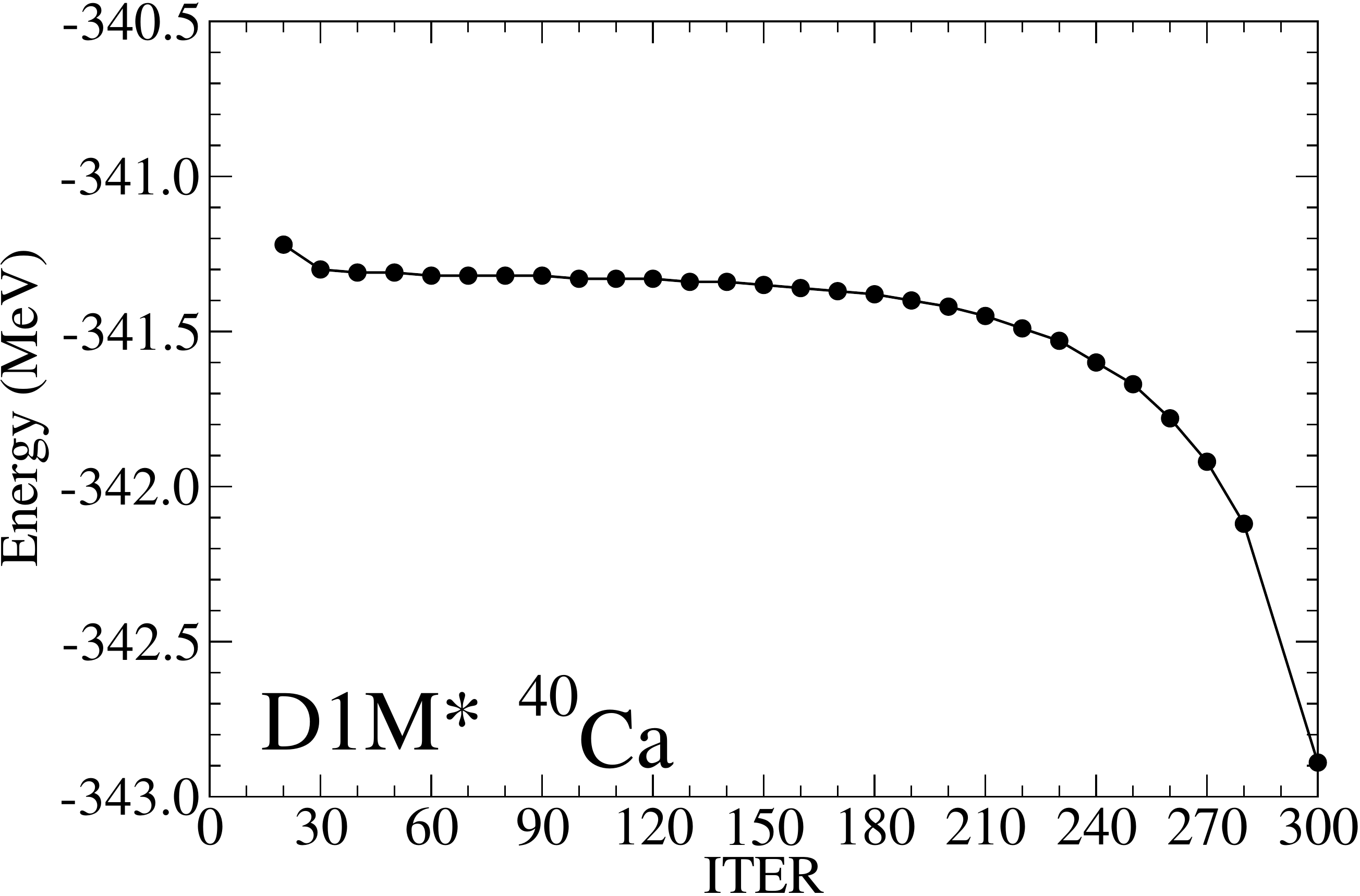}
  \caption{Total energy against the number of iterations of the calculation on a mesh, for $^{40}$Ca with the D1M$^{*}$ interaction.}\label{fig:40Ca_Energy}
\end{figure}

\begin{figure}[h]
 \centering
 \includegraphics[width=0.95\linewidth, clip=true]{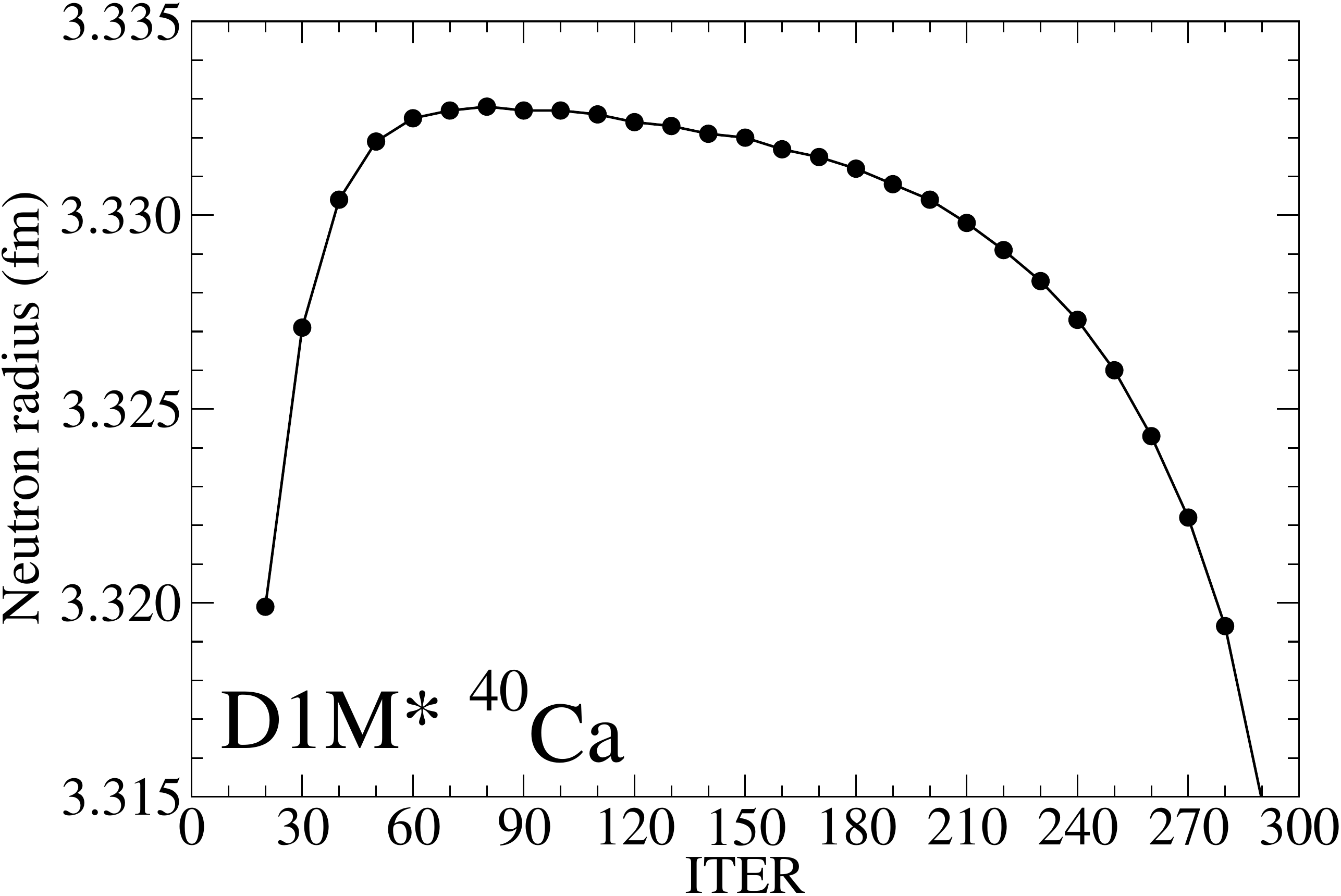}
  \caption{Neutron radius against the number of iterations of the calculation on a mesh, for $^{40}$Ca with the D1M$^{*}$ interaction.}\label{fig:40Ca_rn}
\end{figure}

\begin{figure}[b!]
 \centering
 \includegraphics[width=0.95\linewidth, clip=true]{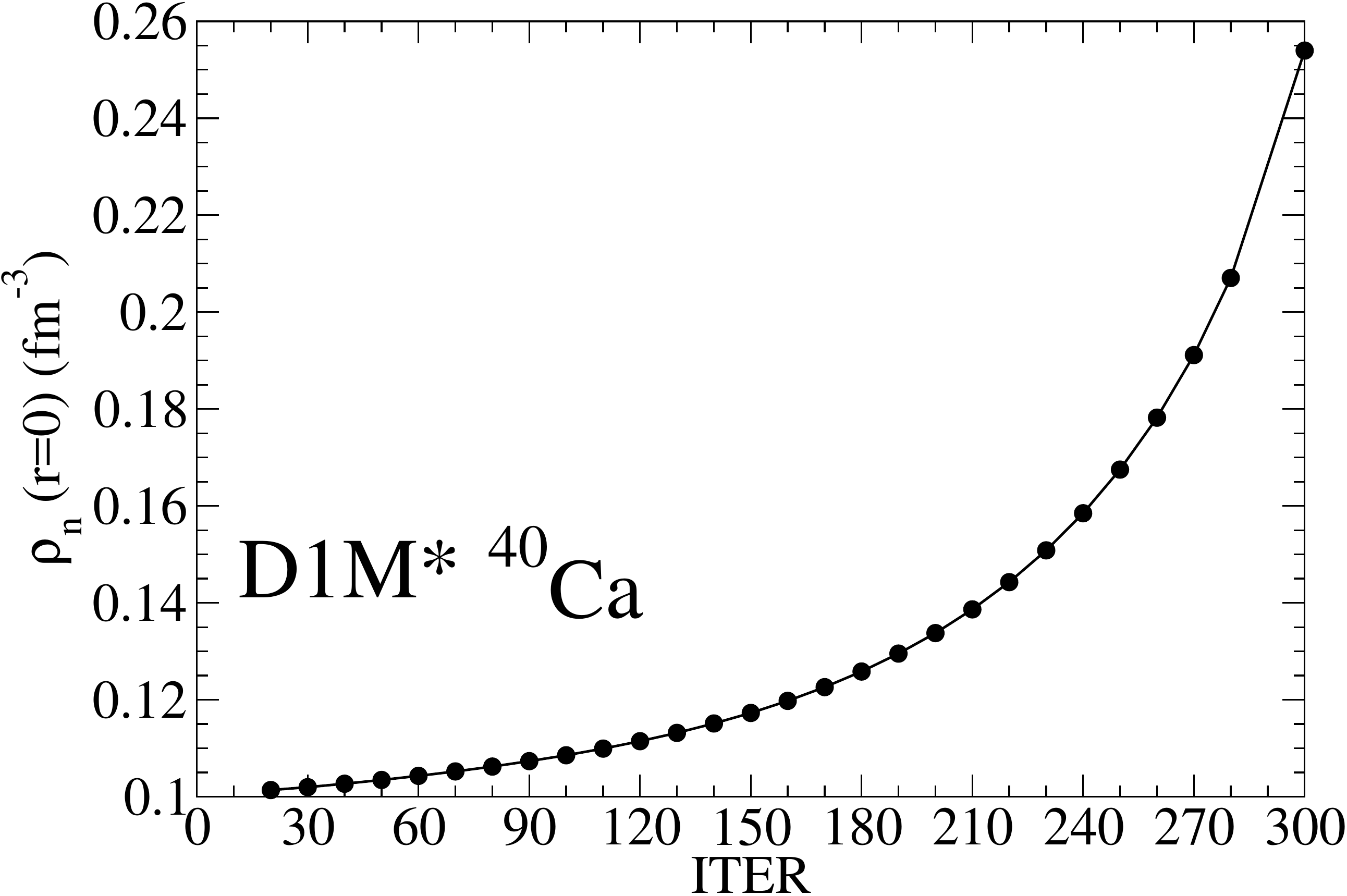}
  \caption{Neutron density at the origin against the number of iterations of the calculation on a mesh, for $^{40}$Ca with the D1M$^{*}$ interaction.}\label{fig:40Ca_rhon_o}
\end{figure}

\begin{figure}[t!]
 \centering
  \vspace{0.8cm}
 \includegraphics[width=0.95\linewidth, clip=true]{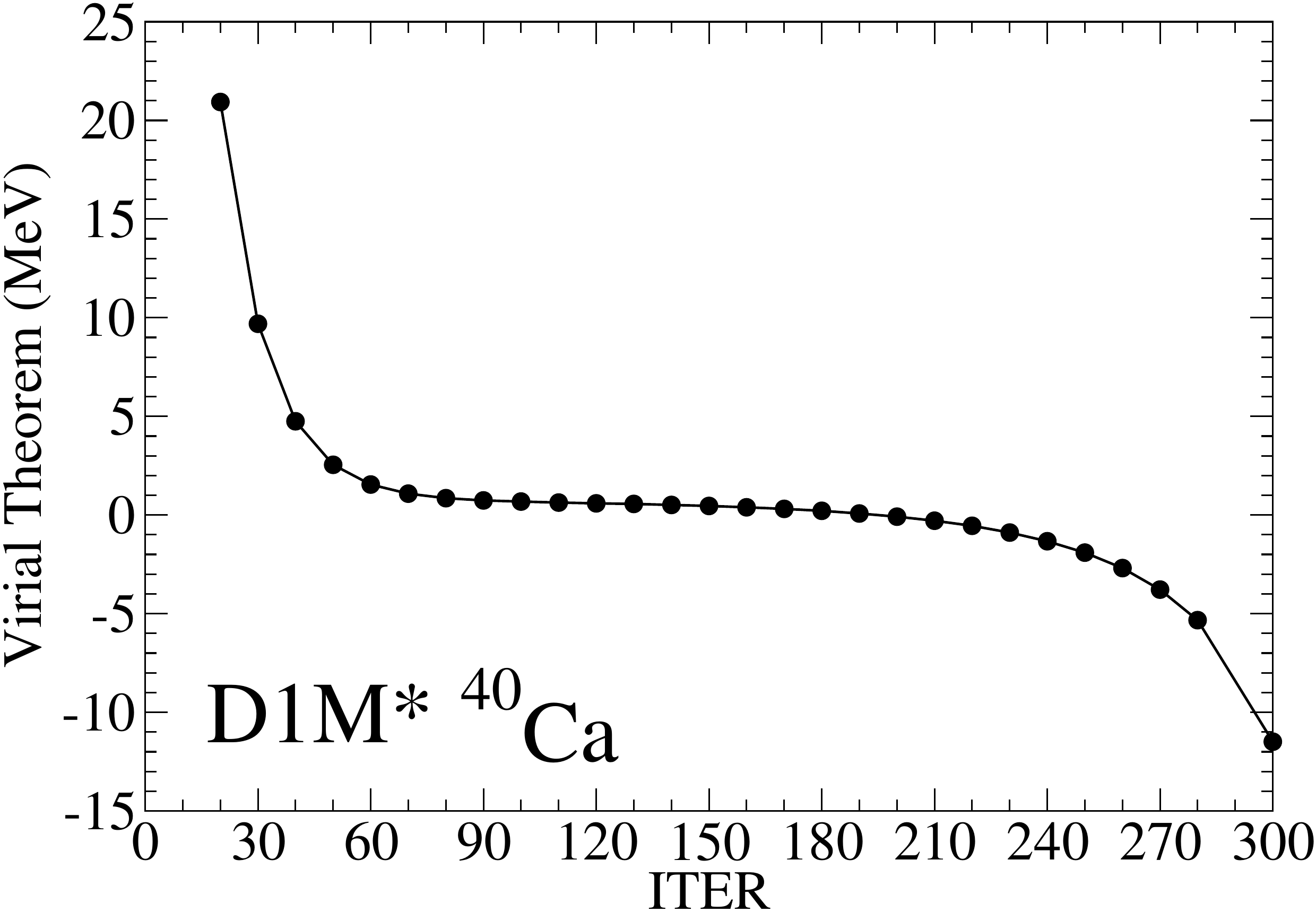}
  \caption{Virial theorem against the number of iterations of the calculation on a mesh, for $^{40}$Ca with the D1M$^{*}$ interaction.}\label{fig:40Ca_Virial}
\end{figure}

\begin{figure}[h!]
 \centering
 \vspace*{0.5cm}
 \includegraphics[width=0.95\linewidth, clip=true]{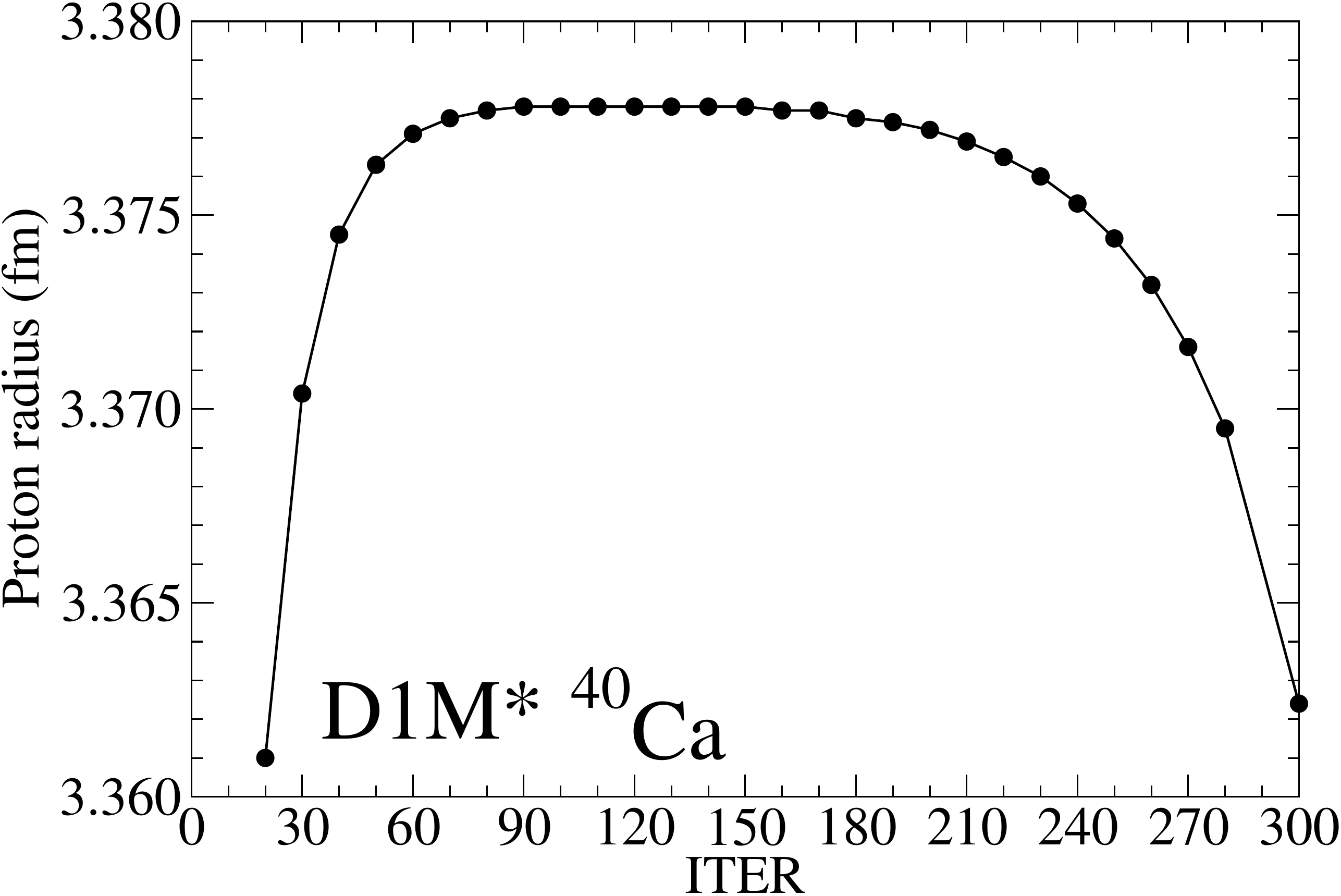}
  \caption{Proton radius against the number of iterations of the calculation on a mesh, for $^{40}$Ca with the D1M$^{*}$ interaction.}\label{fig:40Ca_rp}
\end{figure}

\begin{figure}[b!]
 \centering
 \vspace*{1.2cm}
 \includegraphics[width=0.95\linewidth, clip=true]{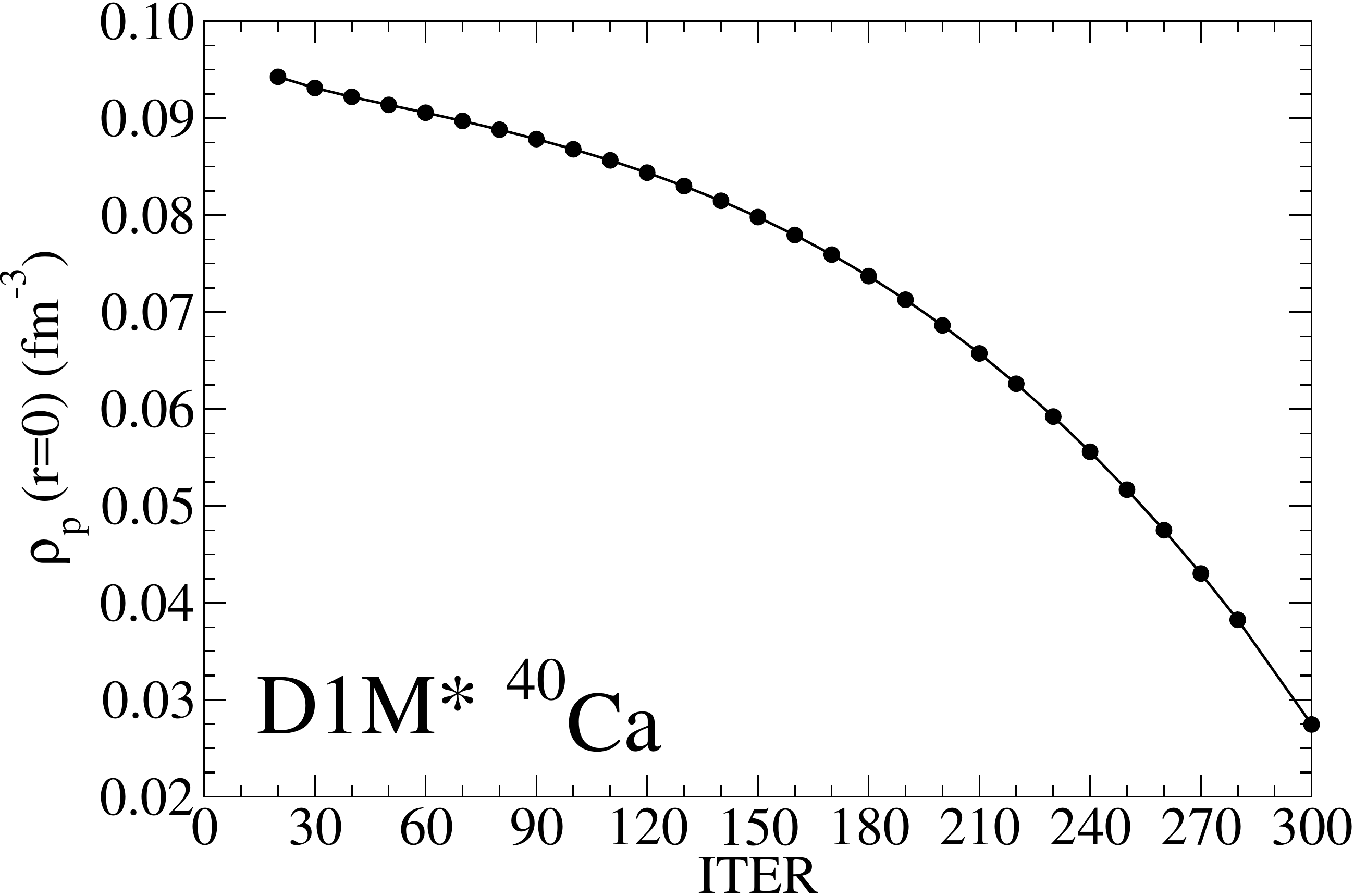}
  \caption{Proton density at the origin against the number of iterations of the calculation on a mesh, for $^{40}$Ca with the D1M$^{*}$ interaction.}\label{fig:40Ca_rhop_o}
\end{figure}

\begin{figure}[t]
 \centering
 \includegraphics[width=1.03\linewidth, clip=true]{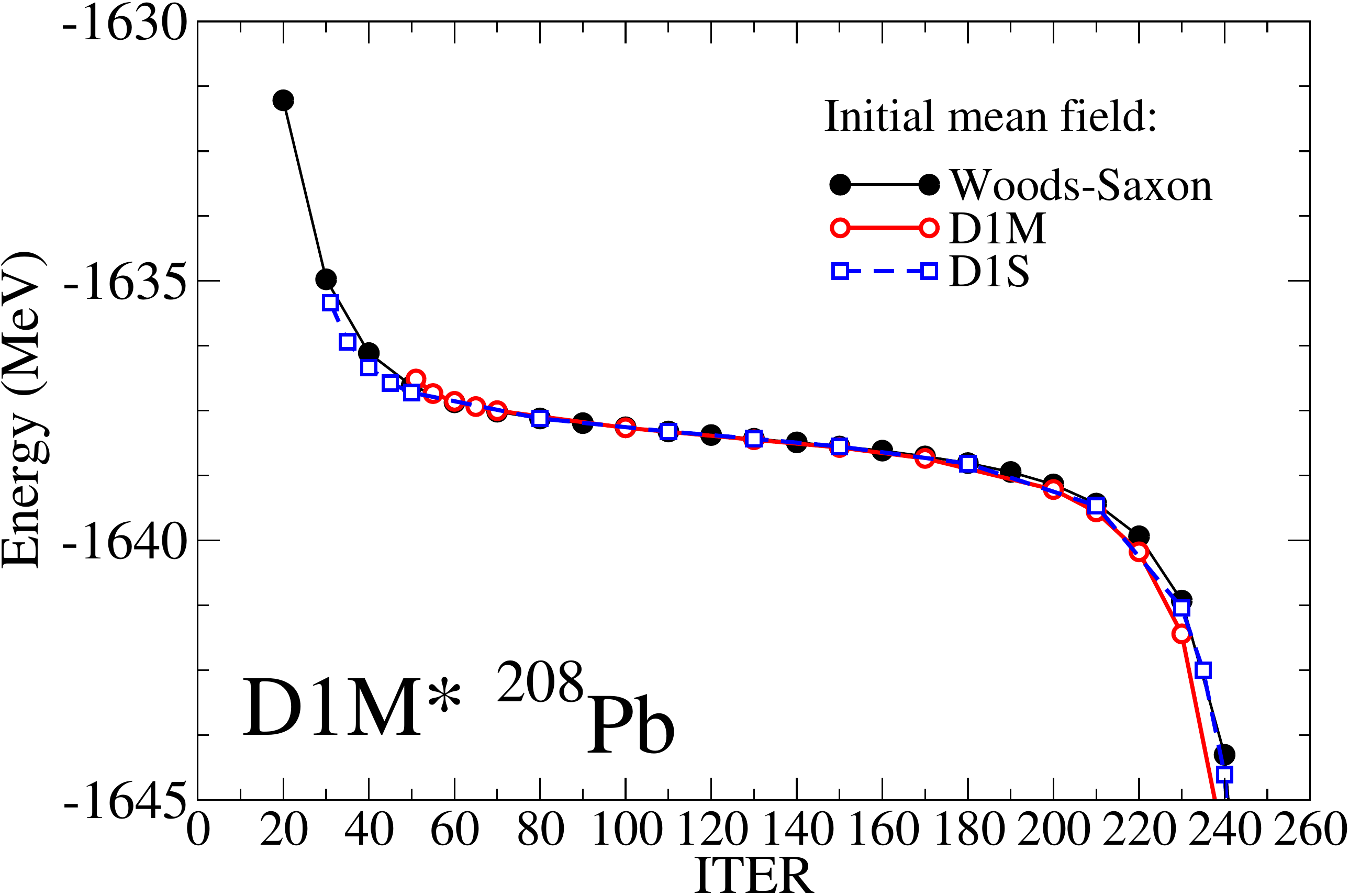}
  \caption{Energy against number of iterations of the calculation on a mesh for $^{208}$Pb with D1M$^{*}$.
  The calculation was started from a WS potential in the black curve (same curve of Fig.~\ref{fig:208Pb_Energy})
  and from the mean field of a converged calculation with the D1M or D1S interactions 
  in the red and blue curves, respectively. The number of iterations on the abscissae corresponds to the case started from 
  a WS potential; the other curves are shifted a few iterations to the right to
  visualize the coincidence of the results in the plateau region.}\label{fig:208Pb_alt_init_pot}
\end{figure}

\begin{figure}[t]
 \centering
 \vspace*{0.1cm}
 \includegraphics[width=1.03\linewidth, clip=true]{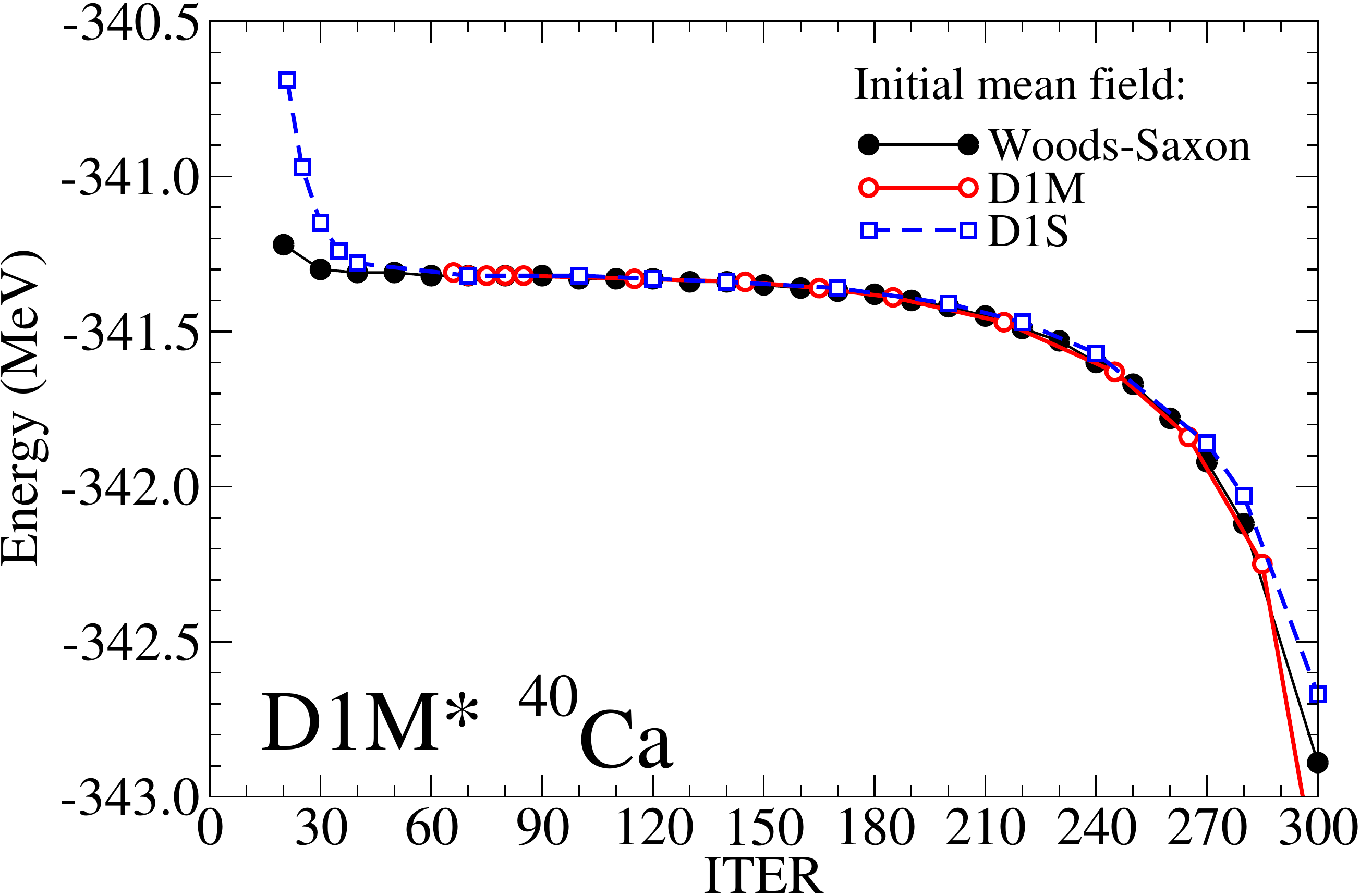}
  \caption{The same as in Fig.~\ref{fig:208Pb_alt_init_pot} but for $^{40}$Ca.}\label{fig:40Ca_alt_init_pot}
\end{figure}

\begin{figure}[t]
 \centering
 \includegraphics[width=1.\linewidth, clip=true]{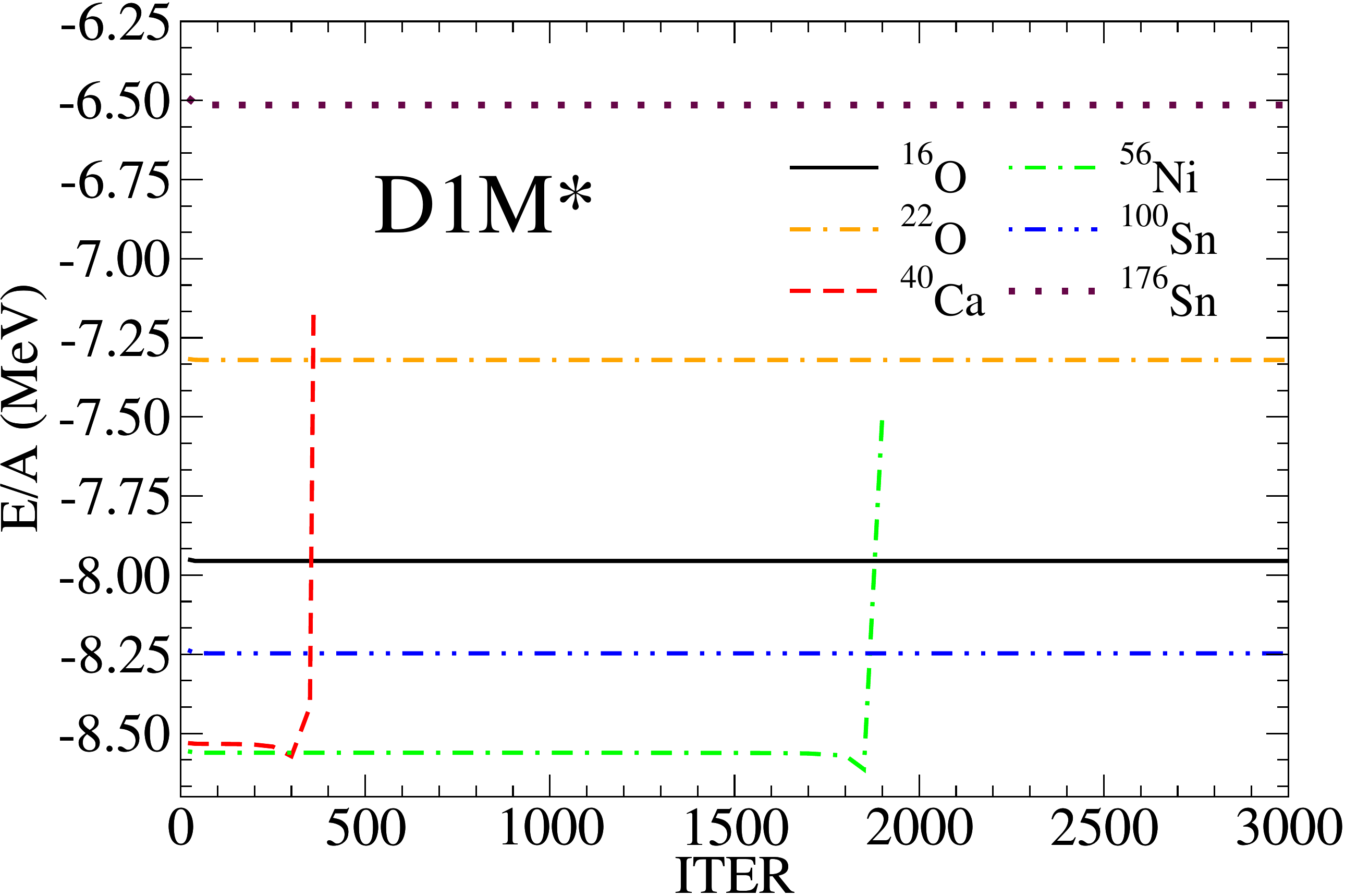}
  \caption{Energy per particle against the number of iterations of the calculation on a mesh, 
  for the $^{16}$O, $^{22}$O, $^{40}$Ca, $^{56}$Ni, $^{100}$Sn and $^{176}$Sn nuclei with the D1M$^{*}$ interaction.}\label{fig:Epart_con_nuclis}
\end{figure}

\begin{figure}[t]
 \centering
% \vspace*{0.8cm}
 \includegraphics[width=1.\linewidth, clip=true]{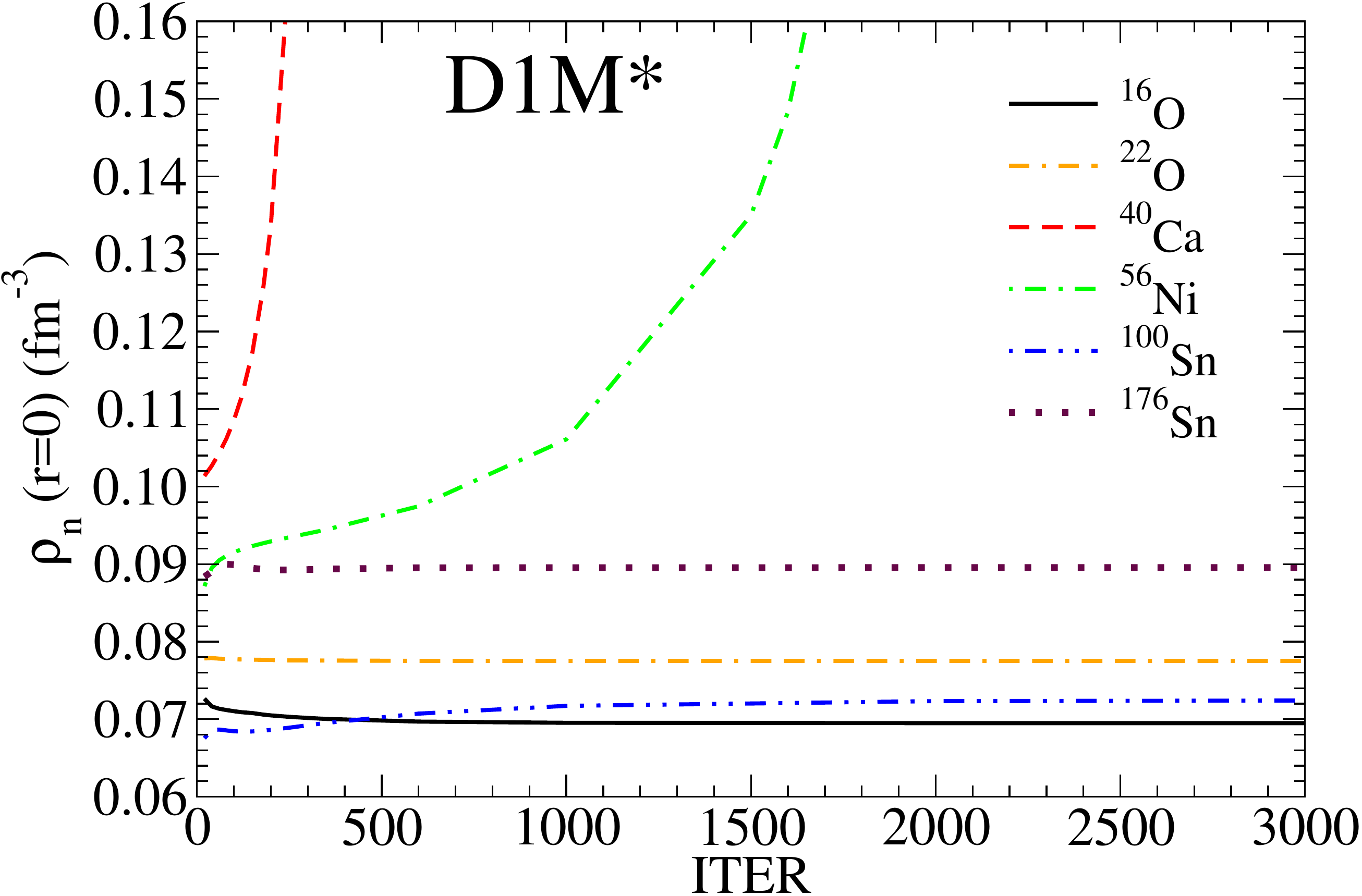}
  \caption{The same as in Fig.~\ref{fig:Epart_con_nuclis} but for the neutron density at the origin.}\label{fig:rhon_con_nuclis}
\end{figure}

\begin{figure}[t]
 \centering
 \includegraphics[width=1.\linewidth, clip=true]{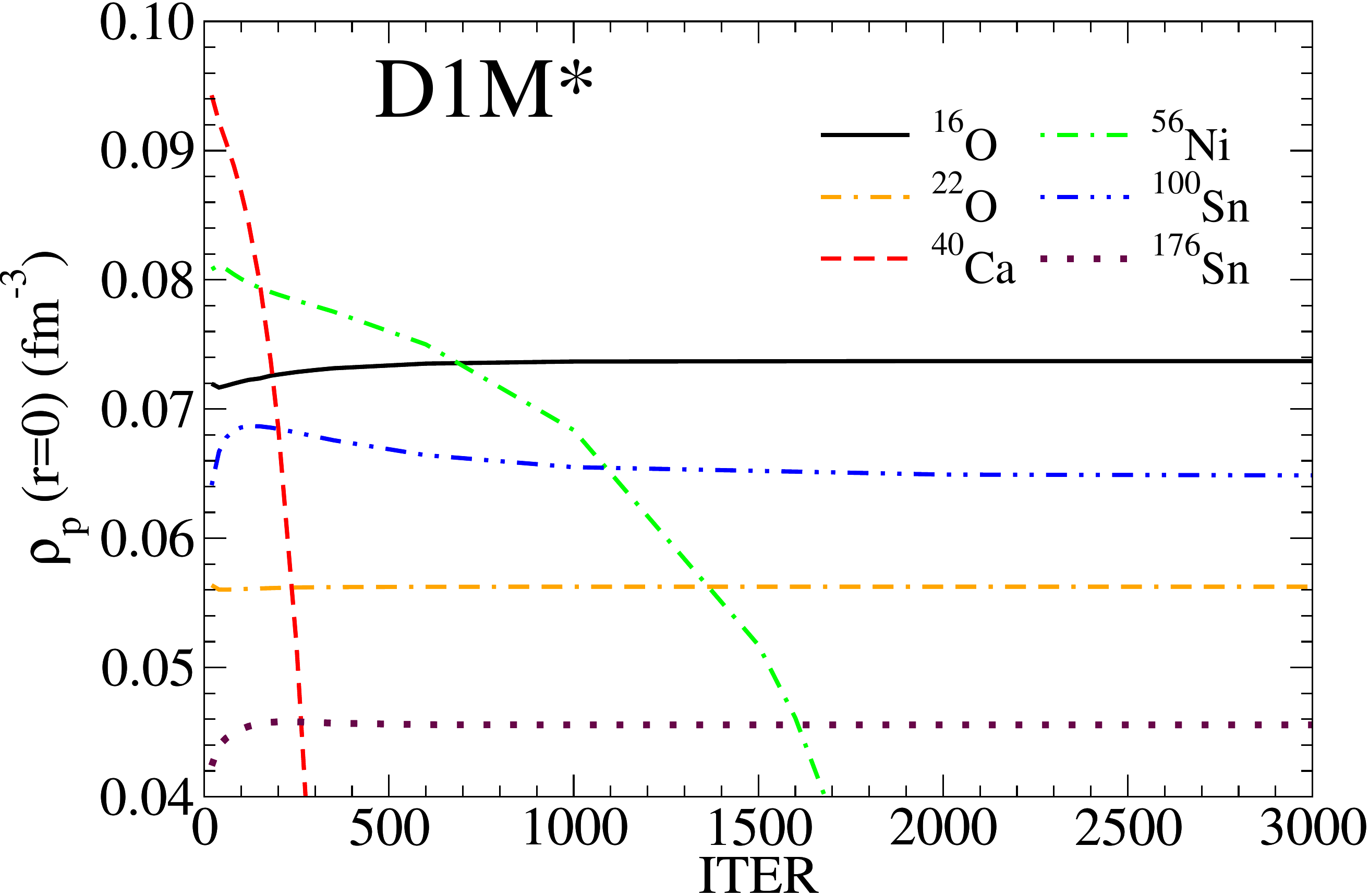}
  \caption{The same as in Fig.~\ref{fig:Epart_con_nuclis} but for the proton density at the origin.}\label{fig:rhop_con_nuclis}
\end{figure}

Additional results of our numerical investigations for D1M*
are collected in Figs.~\ref{fig:208Pb_Energy}--\ref{fig:rhop_con_nuclis}.
We find that the D1M* HF results calculated with the quasilocal approximation show the same instabilities
as the HFB mesh calculations of \cite{martini18}. Our HF calculations of  
the nuclei $^{40}$Ca and $^{208}$Pb with D1M* show a non-convergent behavior as a function of the number
of iterations as can be seen in Figs.~\ref{fig:208Pb_Energy}--\ref{fig:40Ca_rhop_o}. Again, as can be seen 
in Figs.~\ref{fig:208Pb_rhon_ol} and \ref{fig:208Pb_rhop_o} for $^{208}$Pb and  
Figs.~\ref{fig:40Ca_rhon_o} and \ref{fig:40Ca_rhop_o} for $^{40}$Ca, the central densities are 
the most strongly affected quantities by the finite-size instabilities, increasing or decreasing their values  
with the number of iterations, in agreement with the results  
reported in \cite{martini18}. The impact of the finite-size  
instabilities on the energies is shown in Figs.~\ref{fig:208Pb_Energy} and 
\ref{fig:208Pb_Virial} for $^{208}$Pb and Figs.~\ref{fig:40Ca_Energy} and \ref{fig:40Ca_Virial} for $^{40}$Ca.
We note that as a function of the number of iterations the total energies
show a plateau structure, before eventually diverging for a larger number of iterations.
The number of iterations where the energy is nearly stable depends, of course, on
the initialization conditions and the mixing factor. 
This factor combines some quantities obtained in a given iteration with the same 
quantities of the previous iteration in order to slow down and stabilize the  
iterative process. The results in Figs.~\ref{fig:208Pb_Energy}--\ref{fig:40Ca_rhop_o}
were obtained with a mixing factor of 0.9 starting from a Woods-Saxon (WS) shape for the 
initial mean-field potential.
If the mixing factor is changed or the initial potential is changed (e.g., by using 
other WS parameters or the converged mean field of a calculation with D1M or D1S), 
basically the same results are reached in the plateau region (see Figs.~\ref{fig:208Pb_alt_init_pot} 
and \ref{fig:40Ca_alt_init_pot}), although the beginning and width 
of the plateau may involve another number of iterations. An additional 
gauge of the convergence is provided by the stationarity condition arising from the scaling
transformation ${\bf r}\to \lambda {\bf r}$ \cite {bohigas79}, which is equivalent to the  
virial theorem \cite{merzbacher61}. Notice that the stability against  
the scaling transformation is a very strict condition as it comes from the
cancellation of quantities that individually are very large. We display in  
Figs.~\ref{fig:208Pb_Virial} and \ref{fig:40Ca_Virial}
the derivative of the scaled energy with respect to the scaling parameter 
at $\lambda=1$, i.e., $dE(\lambda)/d \lambda|_{\lambda=1}$. We see that 
in a relatively large range of iterations, which coincides with the  
range where the energies are almost flat, 
$dE(\lambda)/d \lambda|_{\lambda=1}$ takes a rather constant value close to zero,
pointing out the near stability of the energy within this number of iterations. 
We have also checked the impact of the instabilities on other quantities directly 
related to the nucleon densities such as the neutron and proton rms radii. We display the rms radii
in Figs.~\ref{fig:208Pb_rn} and \ref{fig:208Pb_rp} (Figs.~\ref{fig:40Ca_rn} and \ref{fig:40Ca_rp})
for $^{208}$Pb ($^{40}$Ca). We see that the impact of the 
instability is relatively more important in $^{208}$Pb than in $^{40}$Ca,  
although these radii can be estimated with less than 0.5\% variation in the  
range of iterations where the energy is quasi stable.  

Further information about the D1M* HF calculations in coordinate space can be extracted from 
Figs.~\ref{fig:Epart_con_nuclis}--\ref{fig:rhop_con_nuclis} displaying the energy per particle 
and the neutron and proton central densities of the nuclei $^{16}$O, $^{22}$O, $^{40}$Ca,  
$^{56}$Ni, $^{100}$Sn and $^{176}$Sn as functions  
of the number of iterations (mixing factor 0.9 in these 
calculations). From these figures we see that $^{40}$Ca becomes  
unstable rather quickly, the energy of $^{56}$Ni is stable until around 1700 iterations, and 
$^{16}$O, $^{22}$O, $^{100}$Sn and $^{176}$Sn are stable (we ran up to 10000 iterations). 
These results and the shell structure of these nuclei suggest that the appearance of 
instabilities in coordinate space is directly related with the existence 
of $s$-orbitals close to the Fermi level. As pointed out in 
\cite{martini18}, and confirmed also in our calculations, the $^4$He nucleus is unstable with D1M* on a mesh. 
Both this nucleus and $^{40}$Ca have an $s$-orbital in the last occupied shell and very 
close to the Fermi level. In the case of $^{56}$Ni the energy gap between the 2$s_{1/2}$ level 
and the Fermi level ($\sim$7 MeV) is larger than in $^{40}$Ca ($\sim$1.5 MeV) but smaller 
than in the $^{16}$O nucleus ($\sim$20 MeV) and the $^{100}$Sn nucleus ($\sim$16~MeV) that are stable. 
In the case of $^{176}$Sn, in the proton spectrum the 2$s_{1/2}$ level remains at $\sim$12 MeV from the Fermi level,
and in the neutron spectrum the 3$s_{1/2}$ level is also deep, at $\sim$10 MeV from the last occupied orbit.

Our findings are in harmony with the claim of Ref.~\cite{martini18} that instabilities 
appear in HF calculations in coordinate space for densities larger 
than about 0.20~fm$^{-3}$ as it happens in the paradigmatic nucleus $^{40}$Ca. 
However, the same $^{40}$Ca nucleus computed without Coulomb, in spite of having a similar value of the central density
as the charged $^{40}$Ca, is stable in the D1M* mesh calculation, as pointed out in \cite{martini18} and
confirmed in our quasilocal approach. 
Hence, some neutron-proton asymmetry is needed to trigger the instability in the calculations 
in coordinate space when $s$-orbitals close to the Fermi level are occupied. 
The role of the $s$-orbitals of the nucleus on the instabilities in a mesh seems to be 
further supported by the case of the $^{22}$O and $^{24}$O isotopes. 
We find in our HF mesh quasilocal calculations that $^{22}$O is stable (the 2$s_{1/2}$ neutron orbital 
is empty) and $^{24}$O is unstable (the 2$s_{1/2}$ neutron orbital becomes the Fermi level). 

\begin{figure*}[h!]\centering
\includegraphics[width=0.7\textwidth,clip=true]{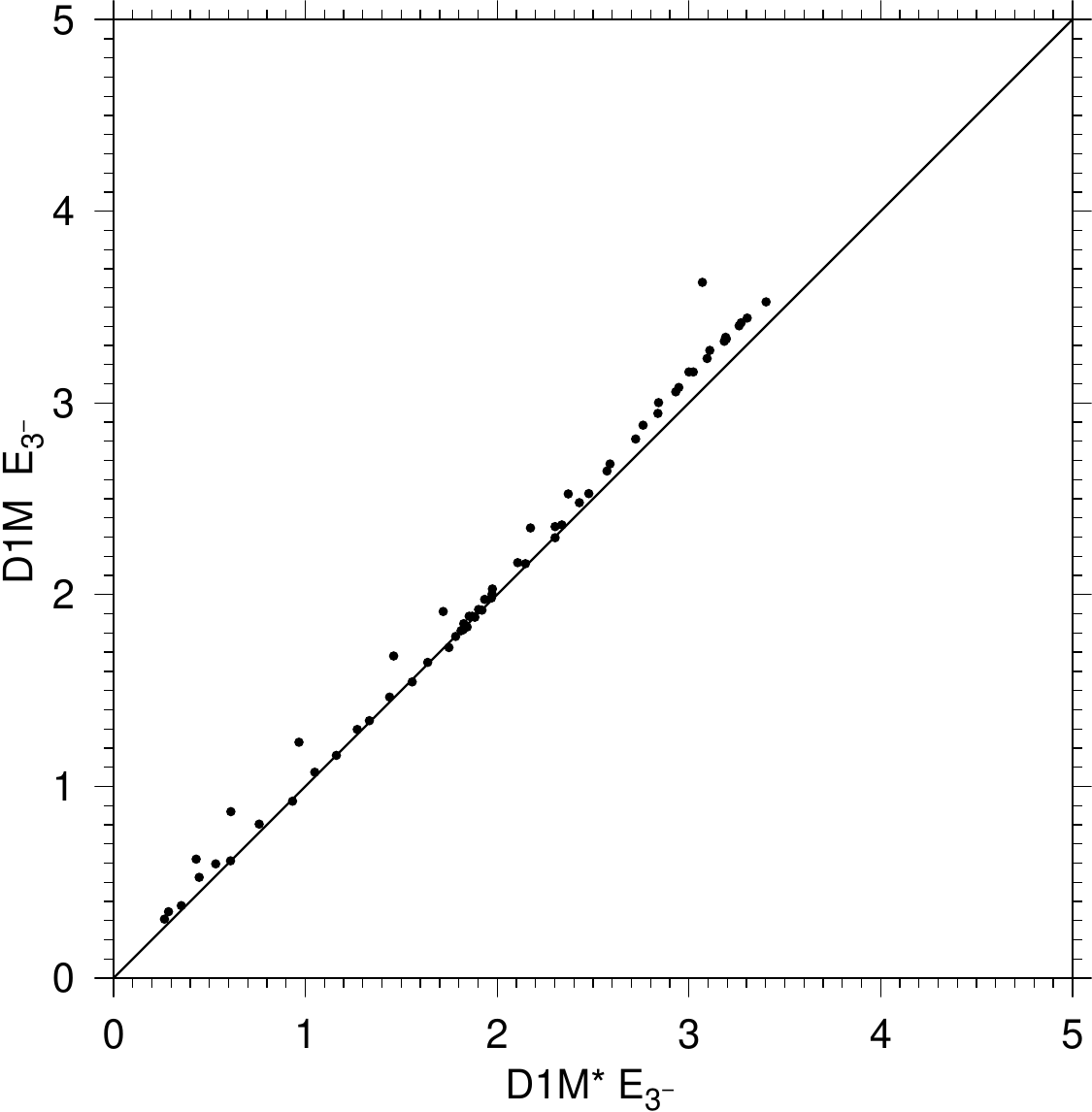}
\caption{Comparison between the excitation energy of the $3^{-}$ state
obtained in a GCM calculation using the octupole degree of freedom
as collective coordinate for both D1M and D1M*.}\label{fig:compare}
\end{figure*}

In the conclusions of Ref.~\cite{martini18} it was argued that beyond mean-field 
calculations with the D1M* interaction may be unsuitable. Although this may
concern RPA calculations, to test the performance of D1M* in a beyond mean-field scenario, 
we have carried out Generator-Coordinate-Method
(GCM) calculations with D1M* and D1M using the octupole moment as collective coordinate for 
neutron-rich Ra and Th isotopes. Again the D1M* and D1M results are very similar
and only show some deviations for shape-coexistent nuclei or those with
very soft potential energy surfaces where tiny changes in the interaction
can lead to some noticeable changes in the observables. As an example of the results,
we present in Fig.~\ref{fig:compare} the plot of the excitation energies
of the $3^{-}$ states obtained in the GCM calculation for both D1M* and
D1M. A perfect agreement between the two forces will lead to all the
points sitting on top of the straight line. In the plot we observe
small deviations for most of the 64 nuclei analyzed, and, in those cases
where the deviations are not negligible, we have checked they correspond
to nuclei with shallow minima where little details are more relevant.

From the discussions above we can conclude that:
\begin{itemize}
	\item We are able to reproduce the results of \cite{martini18} in
	      coordinate space calculations on a mesh using the HF method applied to a 
          quasilocal reduction of the energy density functional associated to the 
          Gogny forces.
	\item In the mesh calculation the density at the origin is the 
          most sensitive quantity to the finite size instability. Other quantities, 
          like the binding energy, radii or the virial theorem reach a broad plateau 
	      with very reasonable values (as compared to stable parametrizations)
	      but those quantities eventually develop an instability if a sufficiently
	      large number of iterations is used.
	\item The instabilities seem to be nucleus dependent and are apparently
	      connected with the relative position of the last occupied $s$-orbital
	      with respect to the Fermi level. 
	\item The HO basis, with its natural ultraviolet cutoff, serves as a 
	      regulator of the finite size instabilities and in all the calculations
	      carried out so far in a large variety of situations no instability is
          found. The results are always comparable to the ones of	stable 
          parametrizations. 
	\item The conclusions of \cite{martini18} regarding the unsuitability
	      of the HO basis for finite nuclei calculations seem to us a bit far-fetched.
	      Although in the limit of infinite HO shells the instabilities 
	      may be present, in the standard framework with Gogny forces where 20 harmonic 
	      oscillator shells are used at most, the convergence of most of the
	      quantities with basis size is very good and there is not a spurious
	      dependence with oscillator lengths.
	\item Finally, we find that D1M* is suitable for beyond mean-field calculations 
	      such as the GCM calculations discussed before.	      
\end{itemize}

We are now in the process of writing a paper showing the
predictions of D1M* for a large set of finite nuclei observables and comparing them
with the ones of D1M. We observe a very good agreement between the 
predictions of the D1M and D1M* parametrizations for finite nuclei (as it should be, given 
the guiding principles in the fit of D1M*, see \cite{gonzalez18}), in all the 
quantities analyzed, which include moments of inertia, spectra of odd nuclei, 
potential energy surfaces as a function of quadrupole and/or octupole multipole 
parameters, $(\beta$, $\gamma)$ planes as required by the 5DCH, and even 
fission barriers in the actinides and the super-heavies.
All these calculations have been carried out using a  HO basis 
with a size as large as possible. We are confident on the
regularization properties of the HO basis to yield good results in finite
nuclei both at the mean-field and beyond mean-field levels even in cases where 
finite-size instabilities may appear in HF calculations on a mesh.

{\small 
%\section*{References}
%\bibliography{mybibfile.bib}

}
\clearpage


\clearpage
\section*{$^{208}$Pb}

\clearpage

\section*{$^{40}$Ca}


%-----------------------------------------------------------------------
\clearpage

\hspace*{5cm}

\pagebreak

\hspace*{5cm}

%-----------------------------------------------------------------------
\clearpage

\clearpage

\end{document}